\documentclass[journal]{IEEEtran}
\usepackage{etex}
\usepackage{setspace}
\usepackage{enumerate}
\usepackage{amsthm,amsmath, amsfonts, amssymb, epsfig, pstricks,pst-node,pst-tree,graphicx,graphics}
\usepackage{./aoslay}
\usepackage{footnote}
\usepackage{enumerate}
\usepackage{flushend,bbm}
\def\setof#1#2{\{\,{#1}\,:\, {#2}\,\}}
\newcommand{\poly}{\,\text{poly}\,}
\newcommand{\Y}{{\bf Y}}
\newcommand{\kn}{{k_n}}
\newcommand{\ptq}{p_{{}_{\mathcal{T},q}}}

\newcommand{\ptildettildeq}{p_{{}_{{\tilde{\mathcal{T}}},{\tilde q}}}}
\newcommand{\T}{\mathcal{T}}
\newcommand{\Ttilde}{{\tilde{\mathcal{T}}}}

\newcommand{\A}{\mathcal{A}}

\newcommand{\E}{{\mathbb E}}
\newcommand{\iid}{\emph{i.i.d.}\xspace}
\newcommand{\indctr}{\mathbbm{1}}
\newcommand{\suffixof}{\preceq}

\newcommand{\w}{{\bf w}}
\newcommand{\s}{{\bf s}}
\newcommand{\U}{{\bf u}}
\newcommand{\V}{{\bf v}}

\providecommand{\Gtilde}{{\tilde G}}

\newcommand{\ctxs}[1]{\textbf{c}_{{#1}}}
\newcommand{\ctxt}{\textbf{c}_\mathcal{T}}
\newcommand{\ctxtt}{\textbf{c}_\mathcal{\tilde{T}}}
\usepackage{tikz}
\usepackage{pgf}
\usepackage{stmaryrd}
\usetikzlibrary{arrows,automata}
\usepackage{mathtools}
\usetikzlibrary{arrows}

\newcommand{\cplngg}{\eta_{{}_{\tilde{G}}}}
\newcommand{\trp}{Q}

\newcommand{\napproxs}[1]{{\,\stackrel{\mathit{{#1}}}{\not\approx}\,}}
\newcommand{\napproxkn}{\napproxs{\kn}}
\newcommand{\approxs}[1]{{\,\stackrel{\mathit{{#1}}}{\approx}\,}}
\newcommand{\approxkn}{\approxs{\kn}}

\newcommand{\prob}{P}
\newcommand{\stn}{\pi}	
\usepackage{tikz}
\usepackage{pgf}
\newcommand{\x}{{\bf x}} 
\newcommand{\ntilde}{{\tilde n}}

\newcommand{\pctxtree}{p_c}

\newcommand{\Yo}{{Y^{'}_j}}
\newcommand{\Zo}{Z^{'}}
\newcommand{\Yos}[1]{{Y^{'}_{j#1}}}
\newcommand{\Yoss}[2]{{Y^{'}_{#1#2}}}

\newcommand{\sYo}{{\sets{\Yos{i}}_{i\ge1}}}
\newcommand{\sZo}{{\sets{\Zo_i}}}

\newcommand{\Yt}{{Y^{''}_j}}
\newcommand{\Zt}{Z^{''}}
\newcommand{\Yts}[1]{{Y^{''}_{j#1}}}
\newcommand{\Ytss}[2]{{Y^{''}_{#1#2}}}
\newcommand{\sYt}{{\sets{\Yts{i}}_{i\ge1}}}

\newcommand{\sZt}{{\sets{\Zt_i}}}

\newcommand{\entrT}{\mathcal{H}_{\T}}
\newcommand{\entrTt}{\mathcal{H}_{\tilde{\T}}} 
\newcommand{\hs}{\mathcal{H}_{\s}}
\newcommand{\hw}{\mathcal{H}_{\w}}

\newcommand{\nsa}{n_{\s a}}
\newcommand{\nwa}{n_{\w a}}

\newcommand{\coplg}{\mu}

\begin{document}
\title{Stationary and Transition Probabilities in Slow Mixing, Long Memory Markov Processes}
\author{Meysam~Asadi,~\IEEEmembership{Student Member,~IEEE,}
        Ramezan~Paravi~Torghabeh,~\IEEEmembership{Student Member,~IEEE,}
        and~Narayana~P.~Santhanam,~\IEEEmembership{Member,~IEEE,}
\thanks{The authors are with the Department of Electrical Engineering, University of Hawai`i, M\={a}noa, Honolulu, HI 96822 USA (Email:$\{$masadi, paravi, nsanthan$\}$@hawaii.edu).}
\thanks{Manuscript received May 3, 2013; revised January 14, 2014; accepted April 19, 2014.}
\thanks{This paper was presented in part at the 2013 IEEE International Symposium on Information Theory.}
\thanks{Copyright~\copyright~2013 IEEE. Personal use of this material is permitted.  However, permission to use this material for any other purposes must be obtained from the IEEE by sending a request to pubs-permissions@ieee.org.}
}

\markboth{IEEE Transactions on information theory,~Vol.~XX, No.~X, XXX~2014}%
{Shell \MakeLowercase{\textit{et al.}}: Bare Demo of IEEEtran.cls for Journals}

\maketitle

\begin{abstract}
  We observe a length-$n$ sample generated by an unknown,
  stationary ergodic Markov process (\emph{model}) over a finite
  alphabet $\cA$. \ignore{Motivated by applications in what are known as
  \emph{backplane channels},}
  Given any string $\w$ of symbols from $\cA$ we want estimates of the
  conditional probability distribution of symbols following $\w$, as well as the stationary probability of
  $\w$.
  Two distinct problems that complicate estimation in this setting are
  (i) long memory, and (ii) \emph{slow mixing} which could happen even
  with only one bit of memory.  

  Any consistent estimator in this setting can only converge pointwise over the
  class of all ergodic Markov models. Namely, given any estimator and any sample
  size $n$, the underlying model could be such that the estimator performs
  poorly on a sample of size $n$ with high probability. But can we look at a
  length-$n$ sample and identify \emph{if} an estimate is likely to be accurate?

  Since the memory is unknown \emph{a-priori}, a natural approach is to estimate
  a potentially coarser model with memory $\kn=\cO(\log n)$. As $n$ grows,
  pointwise consistent estimates that hold eventually almost surely (e.a.s.) are
  known so long as the scaling of $\kn$ is not superlogarithmic in $n$. Here,
  rather than e.a.s. convergence results, we want the best answers possible with a
  length-$n$ sample. Combining results in universal compression with Aldous'
  coupling arguments, we obtain sufficient conditions on the length-$n$ sample
  (even for slow mixing models) to identify when naive (i) estimates of the
  conditional probabilities and (ii) estimates related to the stationary probabilities
  are accurate; and also bound the deviations of the naive estimates from true
  values.
\end{abstract}


\IEEEpeerreviewmaketitle

\section{Introduction}
\IEEEPARstart{W}{e} explore the question of estimating a stationary ergodic $\cA$-ary
Markov process (\emph{model}) from a length-$n$ sample generated by
it. Ideally, given any string $\w$ of symbols from $\cA$ we want an
estimate of the conditional probability distribution of symbols
following $\w$, as well as the stationary
probability of $\w$. As with PAC-learning setup~\cite{Val84},
estimates should come with an accuracy guarantee which holds with a
certain confidence.

For the simpler, finite alphabet \iid sources, empirical probabilities
estimated from length-$n$ samples are well understood. The
deviation of empirical estimates from true values is
characterized by the Chernoff bound~\cite{Mot95} and generalized by
Hoeffding bounds~\cite{Grim01}. On the other hand, Markov sources
contain additional biases which have to do with the \emph{mixing} of
the source, or how quickly all states of the process
are explored.

Roughly speaking, an aperiodic, ergodic source has mixed (or explored all states
properly) when the empirical counts of states in the sample reflects their
stationary probabilities. Given a source has mixed, it is therefore possible to
estimate the transition probabilities using the stationary
probabilities. Indeed, most estimation (theoretical or in practice) follows this
sequence of logic.  Most theoretical results prove that the empirical counts of
states reflect their stationary probabilities eventually almost surely, and
build on this to obtain transition probabilities.

In this paper, we are interested in the regime when the source has 
not yet mixed. This breaks the usual approach---when the empirical
counts of states do not reflect stationary probabilities, is it
at all possible to estimate the transition probabilities? And when
the counts of states are not near their stationary probabilities, what
do they actually signify? 

\subsection*{Motivation: Information aggregation on the Internet}
In this paper we focus on the theoretical underpinnings of Markov estimation in
slow mixing regime. However, it may be instructive to consider the following problem
that places estimation in the slow mixing setting in a concrete context. 

There is implicit but not well-modeled bias in widely adopted means of obtaining
news and other information on the Internet. 
Rather than one news channel disseminating information over TV or radio as in
times not too long ago, there is a very broad choice today among news sources.
While it is desirable that citizens are exposed to a variety of information
sources that is not what really happens in, say, a political context. Depending
on political persuasion, one starts off with perhaps the conservative Fox News
web portal or maybe the liberal New York Times. These sites would then link to
various blogs, more news web sites, and so on---but perhaps mostly on the same
side of the political spectrum. Even in webpages that do not clearly fall into
either side, probabilities with which links are chosen are reflective of the
user opinions. Common browsing habits are therefore not likely to explore the
diversity of views on the Internet, but rather be confined to sections of the
network and perhaps consider certain opinions or news more than others. Of
course, not all news is polarized---sports scores, for example, are not.

Contrast the above with a browsing model that motivates a different view of the
Internet---the Google PageRank. Here the hypothetical user follows links at
random from the page she is currently on. In addition, the user may jump to a
random page on the Internet with a specified (\emph{reset}) probability. Such a
random walk is fast-mixing~\cite{HK03}, rendering its allied global properties like PageRank
easy to compute. While PageRank's efficacy in search tasks is quite
self-evident, it does not capture opinions of specific users. Both the NYT and
Fox News may have high PageRank but very few in the United States would rate/use
both these portals highly.

Any walk on a graph with a randomized aspect to it naturally defines a Markov
process made of symbols corresponding to the vertices of the graph. To study the
polarization of views on a topic, suppose we represent each page by a
finite-alphabet signature relevant to the topic at hand. For example, at the
very simplest, consider representing each page by a bit that represents the
presence or absence of a particular combination of keywords. Secondly, we make
the modeling assumption that browsing history captures users preferences---namely,
if we were given the sum total of a user's browsing histories, we could obtain
the probabilities with which the user may follow various pages.

We therefore model the problem with a Markov process $p$ defined by a full binary
tree whose leaves are represented by a suffix-free set $\T \subset
\sets{0,1}^*$.  The leaves of $\T$ are the \emph{states} (browsing histories) of
the process and $Y_i$ is the $i'$th page visited. Let $\ctxt(Y_{-\infty}^0)$ be
the longest suffix of $Y_{-\infty}^0$ in $\T$, then
\vspace{-.2cm}
\[
p(Y_1^n|Y_{-\infty}^0) 
= 
\prod_{i=1}^n 
p(Y_i|\ctxt(Y^{i-1}_{-\infty})).
\vspace{-.2cm}
\]
All users adhere to this general unknown model $p$, but users are distinguished
by modeling their distinct opinions/preferences as the transition probabilities
corresponding to different states (browsing histories) of the process above. It
can be shown (\eg, via the Dobrushin coefficient) that the more polarized user
opinions are, the slower mixing $p$ is. Note that here, it is reasonable to
assume that the incremental information an additional page in the browser
history provides diminishes with the amount of history we already have.

Given a topic, how would one describe the polarization of opinion or information
of that topic? In our formalization, quantifying polarization amounts to obtaining
the transition and stationary probabilities of the process $p$ above. To elucidate,
we ask two questions.

\underline{\textit{What user profiles can we tell apart?}}  We do not know
$\T$---the browsing contexts that fix user preferences and click
probabilities. So, with a finite amount of browsing data, the best we can ask
for is to estimate the click probabilities $\ptq(Y_1|\w)$ for contexts $\w$
where $\w\in\sets{0,1}^\kn$ for some number $\kn$ that depends on the amount $n$
of browsing data we have. As we will see, in the slow mixing case, not all
contexts that appear in the browsing data may be amenable to modeling these
probabilities.

\underline{\textit{What is the global picture?}} The stationary probabilities of
$\w$ that appear in the browsing data, reflecting how prevalant the
distinguishable user profiles are (how relevant each one is). 

If we were to translate our theoretical approach in this paper to a one line
layman summary of how to quantify polarization, we would ask how much common
information should different users see before they begin to agree. While we have
described the motivation that lies behind the slow mixing formulation, this
paper focuses on the general theoretical and statistical underpinnings of the
two problems above (transition and stationary probabilities of slow mixing
Markov processes) rather than the details of how to aggregate information.

\subsection*{Outline of problem}
We first consider two complications while estimating Markov processes in the 
slow mixing setting.

\paragraph*{Difficulties}
 \ignore{No matter what the source is or how the
sample looks like, there are universal algorithms describing the sample using at
most $\cO(\log n)$\footnote{A function $f_n=\cO(g_n)$ if $\exists n_0 \in
  \mathbb{N}$ and $\exists M>0$ such that $f_n\leq M g_n$ for $n\geq n_0$.} bits
more than if the source were known.} 

The first complication is that irrespective of how large the size of the sample
at hand, $n$, is we may not be in a position to reliably provide estimates of the
stationary probabilities.

Consider a length-$n$ sample obtained from the following binary Markov source
with memory one. The transition probability from 1 to 0 in a memory-1
source is $\epsilon\ll 1/n$. By changing the transition probability from 0 to 1
appropriately, we can vary the stationary probabilities of 1s and 0s in a wide
range without changing how a length-$n$ sample will look like.  As specific
examples, consider two binary, one-bit memory Markov sources; the first assigns
the transition probability from 0 to 1 to be $\epsilon$, while the second
assigns 2$\epsilon$. An easy computation (see also Example~\ref{ergodic_ex} in
Section~\ref{longslow}) shows that the stationary probabilities of 1 and 0 are
(1/2,1/2) and (2/3,1/3) respectively.

But, if we start from the context 1, with high probability \emph{both}
sources will yield a sequence of $n$ 1s. We cannot distinguish between
the two sources above with a sample of this size, and therefore it is
futile to estimate stationary probabilities from this sample. This
particular regime where the number of times each state (1 and 0 in
this example) appears do not reflect their stationary probabilities is
often formalized as the \emph{slow mixing} case, see~\cite{LPW09}.

The second complication is that no matter what the sample size
$n$ is, with high probability the set of all strings $\w$ (of any fixed length)
in a length-$n$ sample may have arbitrarily small mass under the stationary
distribution.

To see this, observe that a memory-1 binary source that transitions from 1 to 0
with probability $1-\epsilon$ and 0 to 1 with probability $\epsilon/m$ has the
stationary probability of 1 to be $1/(m+1)$ (see Example~\ref{arbit_stat_dist} in
Section~\ref{longslow}).  Yet if $\epsilon\ll 1/n$, starting
from state 1 we see a sequence of $n$ 1s with high probability. By making $m$
large enough, the probability of 1 (and therefore of any sequence of 1s) can be
made arbitrarily small, illustrating the conundrum. 

\paragraph*{Problem}
If the source cannot always be well estimated as above, we would like to give
the best possible answer from the length-$n$ sample---one which may also depend on
how the data looks. Say, for the sake of a concrete example, that we have a
sample $\x_1$, with $n-\log n$ 1s followed by a string of $\log n$ 0s.  Perhaps,
this may have come from a one-bit memory, slow mixing Markov source as in
Example~\ref{ergodic_ex}.  As we saw, it is futile to estimate stationary
probabilities in this case. Contrast this sample with a new sample $\x_2$, also
with $n-\log n$ 1s and $\log n$ 0s, but $\x_2$ has 0s spread uniformly in the
sequence. Unlike with $\x_1$, upon seeing $\x_2$ we may want to conclude that we
have an \iid source with a high probability for 1.

We therefore ask how best to estimate properties of an ergodic, yet potentially
slow mixing Markov process from a sample of size $n$. As the above example
shows, we have an estimation problem where any estimator can only converge
pointwise to the true values, rather than uniformly over the model class.
Rather than e.a.s. guarantees, given a realization of a Markov
process we attempt to provide deviation bounds for transition and stationary
probabilities of substrings seen in the sample.  We insist that our bounds,
while being model dependent as is to be expected, must however be calculated
using only parameters which are well-approximated from the data at hand.

If the Markov source is completely arbitrary, such bounds will essentially be
trivial. Therefore, we make an assumption justified by the motivation
we consider---that the information provided by a symbol $i$ positions in the
past given everything in between diminishes with $i$. However, we do not assume
a-priori knowledge on the depth of context tree of the process, nor do we assume
that the conditional probabilities given the pasts are bounded uniformly away
from zero.

\subsection*{Results} 
We provide a short background on Markov processes in Section~\ref{Sec2}, while
Section~\ref{s:rsl} contains a formal summary of results.

At a high level, these results show how to look
at a data sample and identify properties of the process that are
amenable to accurate estimation from the sample even if the source is
slow mixing. They also allow us to sometimes (depending on how the
data looks) conclude that certain naive estimators of transition
probabilities (Section \ref{est_ch_par}, Theorem \ref{thm:aggemptht}),
or stationary probabilities (Section \ref{Dev_bound_Naive}, Theorem
\ref{stat-estim}) happen to be accurate \emph{even if the process is
  slow mixing}.

Contrary to most prior work, we first obtain estimates on transition
probabilities. To do so, we use universal compression approaches that do not
require that empirical counts of states be close to their stationary
probability. Interpreting the empirical counts of states from the approximate
transition probabilities is complicated by the fact that stationary
probabilities can be very sensitive functions of the transition
probabilities. What then can we say about empirical counts from the few
approximate transition probabilities obtained from the sample?  We use a
coupling argument~\cite{A83} in Section~\ref{Dev_bound_Naive} to answer this
question.

Finally, since our results do not rely on empirical counts of strings
reaching their stationary probabilities, they could be strengthened
using other arguments in literature in cases where we know that the
counts do reflect stationary probabilities.
\subsection*{Estimation and compression}
In the set of all Markov sources, mixing properties only affect estimation, and
is irrelevant to universal compression~\cite{Fit72,Sht77}. We have already seen
how slow mixing rendered estimation of stationary probabilities impossible in
general if we are only allowed a fixed length sample, no matter how large the
sample size is. But the sequences on which estimation was impossible lend
themselves to good universal compression. 

This must give us a little pause since in the finite alphabet \iid case,
universal compression and estimation go hand in hand. Specifically, we will
compare two cases---(i) the set of all \iid binary models, and (ii) the set of
all binary Markov models with memory one. The first \iid collection is well
compressible and universal compression algorithms with only sublinear redundancy
$\Theta(\log n)$ exist\footnote{A function $f_n=\Theta(g_n)$ if $f_n=\cO(g_n)$
  and $g_n=\cO(f_n).$}. Here, it is also possible to estimate the underlying
distribution using a good universal compressor. Specific examples include the
Krichevsky-Trofimov~\cite{kt81} (also known as the add 1/2 rule) or
Laplace~\cite{Lap:pe,gc94} (add 1 rule) approaches. As a more complex example,
the Good-Turing estimator (see~\cite{Goo53}) can also be interpreted as being
obtained from such a universal description~\cite{OSZ03:agt} in a more general
setting---where data is exchangeable, rather than \iid.

In the second Markov case as well, universal compression algorithms~\cite{WST95}
can compress the data well, again with redundancy that is only sublinear as
$\Theta(\log n)$. But universal compression algorithms cannot be used to always
infer stationary properties of the source, as illustrated in the examples above.
Put another way, while we may not be able to always estimate stationary
properties of sources, we can compress sequences generated by slow mixing
sources well.

\ignore{The particular application we are motivated by arises in high speed chip-to-chip
communications, and is commonly called the \emph{backplane
  channel}~\cite{FJ06}. Here, residual reflections between inter-chip connects
form a significant source of interference. Because of parasitic capacitances,
the channel is highly non-linear as well, and consequently the residual signal
that determines the channel state is not a linear function of past inputs as in
typical interference channels. We therefore abstract the backplane channel with
a model where the output is not necessarily a linear function of the input, and
in addition, the channel encountered by any input symbol is determined by the
prior outputs. The joint input/output sequence of such channel models are then
Markov processes, and estimation of the channels is analogous to estimating the
stationary- and transition-probabilities of Markov processes.  Furthermore,
estimating the information rates supported across such channels turns out to be
equivalent to estimation of entropy rates of Markov processes. See~\cite{APS13}
for more details on channel estimation. To retain focus on the basic
fundamentals of the problem, in this paper we concentrate mainly on estimating
parameters related to Markov processes.
The organization of the paper is as follows: After reviewing prior work in
Section~\ref{PriorWork}, we provide background on Markov processes in
Section~\ref{Sec2}. Then, in Section \ref{longslow}, we outline the difficulties
encountered in estimation and in Section~\ref{s:rsl}, we have summary of
results. In Sections~\ref{naive_est}-\ref{est_ch_par}, we develop results on
transition probabilities and in
Sections~\ref{stopping_times}-\ref{Dev_bound_Naive}, we develop results on the
stationary probabilities.}

\section{Prior Work}
\label{PriorWork}
\subsection{Prior work on compression of Markov processes}
For Markov processes with known memory $k$, optimal redundancy
rates for the universal compression and estimation have been
established, see~\eg~\cite{Kr93} for an overview and
also~\cite{Ry08,Ry88,Ry90}. These universal compression results imply
consistent estimators for \textit{probabilities of
  sequences}. Moreover, the rate of convergence of these estimators
can be bounded uniformly over the entire memory$-k$ Markov model
class~\eg~\cite{ris84,WRF95,WST95,TSW93}. This rate typically depends
exponentially on $k$ and diminishes with the sample length as $\log n/n$.
We point out two complications when confronted with our problem.

First, we deal with the case of unbounded memory---namely no
$a-priori$ bound on $k$. For the set of all finite memory Markov
sources, only weakly universal~\cite{Kie78} compression
schemes---those that convergence in a pointwise sense---can be
built (see~\cite{wil98} for a particularly nice construction). Namely,
the convergence of the weak universal algorithms varies depending on
the true unseen memory of the source. However, as we will see in
Example~\ref{dependency}, it may be impossible to estimate the memory
of the source from a finite length sample. There has been a lot of
work on the topic of estimating the memory of the source consistently
when a prior bound on the memory is unknown,
see~\cite{CT06,G06,GMS08,GL11}---but as one would expect, given a
finite length sample no estimator developed will always have a good
answer.

Second, despite the positive result for estimation of the
\textit{probabilities of sequences}, as mentioned in the introduction
there can not be estimators for transition and stationary
probabilities whose rate of convergence is uniform over the entire
model class. This negative observation follows simply because of the
way we are forced to sample from slow mixing processes---and this is a
complication compression does not encounter. For instance, in the
example outlined above in the introduction, both samples $\x_1$ and
$\x_2$ can be well compressed by universal estimators, but estimation
is a whole different ballgame.

The complications above apart, the nature of questions we ask is
different as well. Rather than consistency results, or establishing
process-dependent rates of convergence for the case where the memory
can be unbounded, we ask how to give the best possible answer with a
given sample of length $n$. It is not to say, however, that the above
compression results are irrelevant to our problem. Far from it, one of
our results, Theorem~\ref{thm:aggemptht} in Section~\ref{est_ch_par}
builds on (among other things) the universal compression results
obtained for $k-$memory processes.

\subsection{Prior work on estimation of Markov processes}
Estimation for Markov processes has been extensively studied and falls
into three major categories (i) consistency of
estimators~\eg~\cite{R83,BW99,CT06,G06,MW07}, (ii) guarantees on
estimates that hold eventually almost surely~\eg~\cite{C02,CS00}, and
(iii) guarantees that hold for all sample sizes but which depend on
both transition and stationary probabilities~\eg~\cite{GMS08,GL11,CT10,GF08}. The list above
is not exhaustive, rather it focuses on the work closest to the
approaches we take.

As mentioned earlier, performance of any estimator cannot not be
bounded uniformly over all Markov models, something reflected in the
line (iii) of research and in our work. While unavoidable, it poses a
problem since the deviation bounds now depend on the unknown
model. How then do we say if our estimate is doing well? Our thrust in
this paper focuses on exactly this question---it is not just about
consistency, rather that we want to gauge from the observed sample
\emph{if} our estimator is doing well relative to the unknown
probability law in force.

In \cite{MW07} a survey on consistent estimators for conditional
probabilities of Markov processes is provided. For instance, given a
realization of a Markov process, they provide a sequence of estimators
for transition probabilities along some of time steps which converges
almost surely to the true values.

Consistent estimators for the order of Markov processes have been
studied in prior literature~\eg\cite{MW05,CS00,VH11,MGZ89}.  In
\cite{CS00,VH11}, the penalized maximum likelihood technique is used
to provide a consistent order estimator.  In \cite{MGZ89}, a different
consistent order estimator based on empirical counts is proposed which
minimizes the asymptotic underestimation exponent while keeping the
overestimation exponent at a certain level. Using the same technique,
in \cite{FCN96} an optimal order estimator is provided, however they
assume a prior upper bound on the memory of underlying process.

In \cite{CT06}, estimation of the minimal context tree of the Markov
process is addressed. Two different information criteria, namely
Bayesian Information Criterion and Minimum Description Length are
used, and consistency of estimation of the underlying context tree was
established provided that the depth of hypothetical trees grow as
$o(\log n).$\footnote{A function $f_n=o(g_n)$ if
  $\lim_{n\to\infty}f_n/g_n=0$.} Moreover, it was shown in \cite{G06}
that when the process has finite memory, the $o(\log n)$ condition is
not necessary for estimation consistency.

In \cite{GMS08,GL11,GF08}, exponential upper bounds on probability of
incorrect estimation of (i) conditional and stationary probabilities
and (ii) the underlying context tree, are provided for variants of
Rissanen's algorithm \textit{context} and penalized maximum likelihood
estimator. The introduced deviation bounds depend on the model
parameters (\eg minimum stationary probability of all contexts
$p_{\min}$, depth of the tree and continuity rate coefficients) of
underlying process.

One particular paper that we would like to highlight is~\cite{CT10},
where the problem of estimating a stationary ergodic process by
finite memory Markov processes based on an $n$-length sample of the
process is addressed.  A measure of distance between the true process
and its estimation is introduced and a convergence rate with respect
to that measure is provided. However, the bounds proved there hold
only when the infimum of conditional probabilities of symbols given
the pasts are bounded away from zero.

In this paper, we see entropy estimation as a means of motivating the
main problems to be posed. The best known results in the extensive
research on entropy rate estimation for Markov processes are again
related to Lempel-Ziv universal lossless data compression
methods~\cite{ZL78}.  See~\cite{MW07} for a survey of other work as well.

\section{Markov processes}
\label{Sec2}
Most notation, while standard, is included for completeness.
\subsection{Alphabet and strings} 
$\mathcal{A}$ is a finite alphabet with cardinality $|\mathcal{A}|$,
$\mathcal{A}^{*}=\bigcup_{k \geq 0} \mathcal{A}^{k}$ and
$\mathcal{A}^{\infty}$ denotes the set of all semi-infinite strings of
symbols in $\mathcal{A}$.

We denote the length of a string $\U=u_1\upto u_l\in\cA^l$ by $|\U|$,
and use $\U_{i}^{j}=(u_{i}, \cdots, u_{j})$. The concatenation of strings $\w$ and $\V$
is denoted by $\w\V$. A string $\V$ is a \textsl{suffix} of $\U$,
denoted by $\V\preceq \U$, if there exists a string $\w$ such that
$\U=\w\V$. A set $\cT$ of strings is \emph{suffix-free} if no string
of $\cT$ is a suffix of any other string in $\cT$.

\subsection{Trees} 
As in~\cite{WST95} for example, we use full $\cA-$ary trees to
represent the states of a Markov process. We denote full trees
$\cT$ as a suffix-free set $\mathcal{T} \subset \mathcal{A}^{*}$ of
strings (the \emph{leaves}) whose lengths satisfy Kraft's lemma with
equality.  \ignore{\textit{tree} if no $\U_1\in \mathcal{T}$ is a
  suffix of any other $\U_2\in \mathcal{T}$.}  The depth of the tree
$\mathcal{T}$ is defined as $\kappa(\mathcal{T})=max\setof{|\U|}{\U \in
  \mathcal{T}}$.  A string $\V\in \mathcal{A}^{*}$ is an
\textit{internal node} of $\mathcal{T}$ if either $\V\in \mathcal{T}$
or there exists $\U\in \mathcal{T}$ such that $\V\preceq \U$.  The
\textit{children} of an internal node $\V$ in $\mathcal{T}$, are those
strings (if any) $a\V, a \in \mathcal{A}$ which are themselves either
internal nodes or leaves in $\mathcal{T}$.

For any internal node $\w$ of a tree $\mathcal{T}$, let
\ignore{$\mathcal{T}_{\w}$ denote the set of leaves in $\mathcal{T}$
  such that $\w$ is the suffix of all of them, i.e., }
$\mathcal{T}_{\w}=\{\U \in \mathcal{T}: \w \preceq \U \}$ be the
subtree rooted at $\w$.  Given two trees $\mathcal{T}_{1}$ and
$\mathcal{T}_{2}$, we say that $\mathcal{T}_{1}$ is included in
$\mathcal{T}_{2}$ ($\mathcal{T}_{1} \preceq \mathcal{T}_{2}$), if all
the leaves in $\mathcal{T}_{1}$ are either leaves or internal nodes of
$\mathcal{T}_{2}$.

\subsection{Models} 
Let ${\cal P}^+(\A)$ be the set of all probability distributions on
$\A$ such that every probability is strictly positive.  

\begin{figure}[!t]
\begin{center}
\includegraphics[trim=5cm 12cm 5cm 11cm, clip,scale=.7]{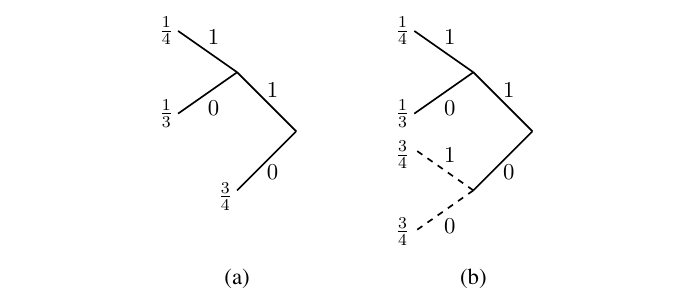}
%
%
%
%
%
\caption{(a) States and parameters of a binary Markov process in Example \ref{property1}, (b) Same Markov process reparameterized to be a complete tree of depth 2. We can similarly reparameterize the process on the left with a complete tree of any depth larger than 2.}
\label{figure:figprop1}
\end{center}
\end{figure}

\bDefinition A context tree $\emph{model}$ is a finite full tree
$\T \subset \A ^*$ with a collection of probability distributions
$q(\cdot|\s)\in {\cal P}^+(\A)$ assigned to each $\s\in \T$. We will refer to
the elements of $\mathcal{T}$ as $\emph{states}$ (or $\emph{contexts}$), and $q(\T) =
\{q(a|\s): \s\in \T, \, a\in \A\}$ as the set of \emph{state transition
probabilities}.
\label{CTX_TREE}
\eDefinition

\indent Every model $(\T, q(\T))$ allows for an irreducible, 
aperiodic\footnote{Irreducible since $q(\cdot|\s)\in\cP^+(\cA)$, aperiodic since any state $\s\in\T$ can
be reached in either $|\s|$ or $|\s|+1$ steps.} and ergodic~\cite{Fel57}
Markov process with a unique stationary distribution $\stn$
satisfying
\begin{equation}
\stn \, \trp=\stn,
\label{STN_DST}
\end{equation}
where $\trp$ is the standard transition probability matrix formed using
$q(\T)$. 
  Let
$\ptq$ be the unique stationary Markov process $\{\ldots
,Y_0,Y_1,Y_2,\ldots\}$ which takes values in $\A$ satisfying
\begin{equation*}
\ptq( Y_1 | Y_{-\infty}^0)  = q(Y_1|\s), 
\label{stationary}
\end{equation*}
where $\s$ is the unique suffix $\s\suffixof Y_{-\infty}^0$ in $\T$ denoted by
$\ctxt(Y_{-\infty}^0)$.  Namely the mapping $\ctxt$ maps any (long enough)
sequence to its unique suffix in $\T$. When the argument of $\ctxt$ is an
internal node of $\T$, we leave the image of the mapping undefined.

As a note, when we write out actual strings in transition
probabilities as in $q(0|1000)$, the state $1000$ is the sequence of
bits as we encounter them when reading the string left to right. If 0
follows $\cdots1100$, the next state is a suffix of $\cdots11000$, and if 1
follows $\cdots1100$, the next state is a suffix of $\cdots11001$. 

\bObservation
\label{obs:indp}
A useful observation is that any model $(\cT,q(\cT))$ yields the same Markov process as a model $(\cT',q(\cT'))$ where
$\cT\preceq \cT'$ and for all $\s'\in \cT'$, $q(\cdot |\s')=q(\cdot|\ctxt(\s'))$.  
\eObservation
\bExample
\label{property1}
Let $(\cT,q(\cT))$ be a binary Markov process with $\T=\{11, 01, 0\}$ and $q(1|11)=\frac{1}{4}, q(1|01)=\frac{1}{3}, q(1|0)=\frac{3}{4}$ as shown in Fig. \ref{figure:figprop1}. (a). Observe that
Fig. \ref{figure:figprop1}. (b) shows the same Markov process as a model $(\cT',q(\cT'))$ with $\T'=\{11, 01, 10, 00\}$  
satisfying conditions in Observation~\ref{obs:indp}.
\eExample

A couple of points about the notation. For any string $\U$, not just
strings in $\cT$, we will use \emph{stationary probability} of $\U$,
$\stn(\U)$, to mean $\ptq(Y_1^{|\U|}=\U)$. If $\s\in\cT$, our notation
is redundant---the transition probability $q(a|\s)$
and $\ptq(a|\s)$ are synonymous. However, if $\U\notin\T$, we will
only use $\ptq(a|\U)$ and avoid using $q(a|\U)$.

\ignore{As emphasized in the introduction, we do not assume the true
model is known nor do we assume it is fast
mixing. We would like to know if we can
estimate the transition and stationary probabilities of
various states even when we are in the domain where the
mixing has not happened.}

\section{Difficulties in Estimation}
\label{longslow}
It is quite possible all strings in a finite sample, no matter how large, have
arbitrarily small mass under the stationary distribution.  We illustrate this in
Example \ref{arbit_stat_dist} below. Our results, particularly Theorem
\ref{stat-estim} incorporates this phenomenon, and we try to provide the best
results despite this apparent difficulty.

\ignore{\indent Notwithstanding the above, there are two other distinct
difficulties in estimating Markov processes as the ones we are
interested in. The first is memory that is too long to handle given
the size of the sample at hand.  The second issue is that even though
the underlying process might be ergodic, the transition probabilities
are so small such that the process effectively acts like a non-ergodic
process given the sample size available. We illustrate these problems
in examples \ref{dependency} and \ref{ergodic_ex}.
In the following example, we show that}

\bExample Let $\A=\{ 0,1\}$ and $\T=\{ 0,1
\}$ with $q(1|1)=1-\epsilon$, and $q(1|0)=\frac{\epsilon}{m}$. For
$\epsilon>0$ and a constant $m\in \mathbb{R}$ with $m>\epsilon$, this
model represents a stationary ergodic Markov process $\ptq$ with stationary
distributions $\stn(1)=\frac{1}{m+1}, \stn(0)=\frac{m}{m+1}$. Note
that $\stn(1)$ can be arbitrarily small for sufficiently large $m$.

Now suppose we have a length-$n$ sample with $\epsilon\ll 1/n$.  If we
start from 1, with high probability we see a sequence of $n$
consecutive 1's. For instance, if $\epsilon=1/n^j$ for some $j\geq 2$, then with probability $\geq 1-1/n$ under $\ptq$, we see a sequence of $n$
consecutive 1's.  Clearly, the stationary probability of any sequence
of 1's is $\leq \frac{1}{m+1}$, and this can be made arbitrarily small
by choosing $m$ large enough.
\label{arbit_stat_dist}
\eExample
\begin{figure}[!b]
\begin{center}
\includegraphics[trim=5cm 12cm 5cm 11cm, clip,scale=.7]{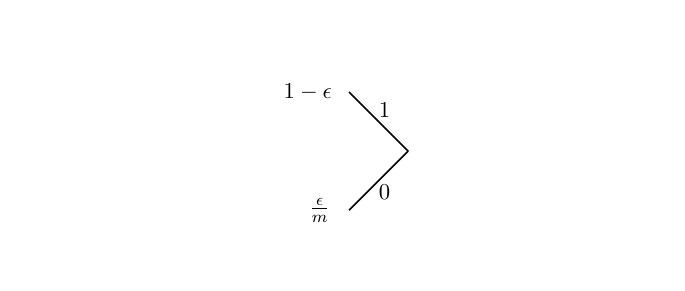}
\caption{Markov processes in Example \ref{arbit_stat_dist} with stationary probabilities $\stn(1)=\frac{1}{m+1}$ and $\stn(0)=\frac{m}{m+1}$.}
\label{ergodic_ex_fig}
\end{center}
\end{figure}

The next example illustrates one pitfall of having no bound on the memory.
We therefore require that dependencies die down by requiring that conditional
probabilities satisfy~\eqref{eq:condition} in Section~\ref{EST_DEV_B}.

\bExample
\label{dependency}
Let $\T=\A^k$ denote a full tree with depth $k$ and $\A=\{0,1\}$.
Assume that $q(1|0^k)=2\epsilon$ and $q(1|10^{k-1})=1-\epsilon$ with
$\epsilon>0$, and let $q(1|\s)=\frac{1}{2}$ (where $0^{k}$ indicates a
string with $k$ consecutive zeros) for all other $\s\in\T$.  Let
$\ptq$ represent the stationary ergodic Markov process associated with
this model. Observe that stationary probability of being in state
$0^{k}$ is $\frac{1}{2^{k+1}-1}$ while all other states have
stationary probability $\frac{2}{2^{k+1}-1}$.  Let $Y_1^n$ be a
realization of this process with initial state $1^k\preceq
Y^0_{-\infty}$.  Suppose $k\gg \omega(\log n)$.  \footnote{A function
  $f_n=\omega(g_n)$ if $\lim_{n\to\infty}f_n/g_n=\infty$.}  With high
probability we will never find a string of $k-1$ zeros among $n$
samples, and every bit is generated with probability 1/2. Thus with
samples of size $n$, no matter how large $n$ may be, with high
probability we cannot distinguish certain long-memory processes from
even an $\iid$ Bernoulli(1/2) process.
\begin{figure}[!t]
\begin{center}
\includegraphics[trim=5cm 9cm 5cm 8cm, clip,scale=.6]{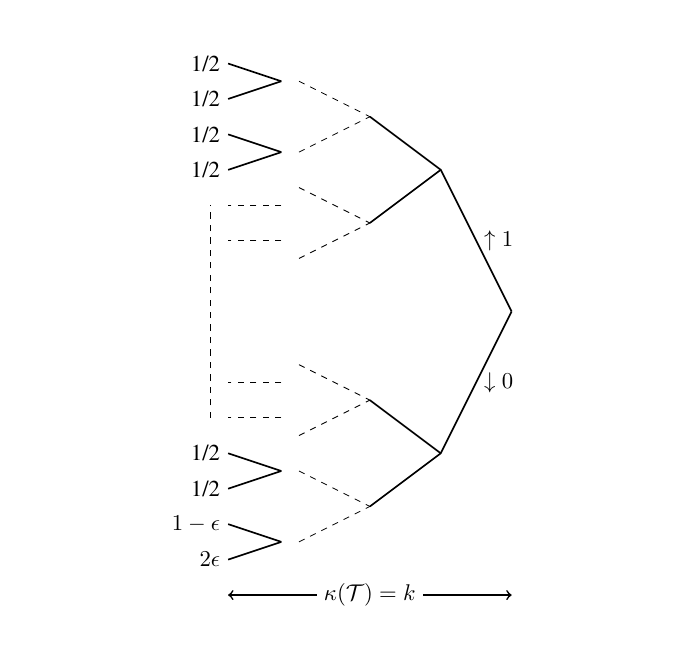}
\caption{Markov process in Example \ref{dependency}. With high probability, we cannot
distinguish $\ptq$ from an $\iid$ Bernoulli(1/2) process if the sample size $n$ satisfies $k\gg \omega(\log n)$.}
\label{dependencyfig}
\end{center}
\end{figure}  \eExample
The third example illustrates complications arising from mixing properties while
estimating stationary probabilities.
\begin{figure}[!b]
\begin{center}
\includegraphics[trim=5cm 12cm 5cm 11cm, clip,scale=.7]{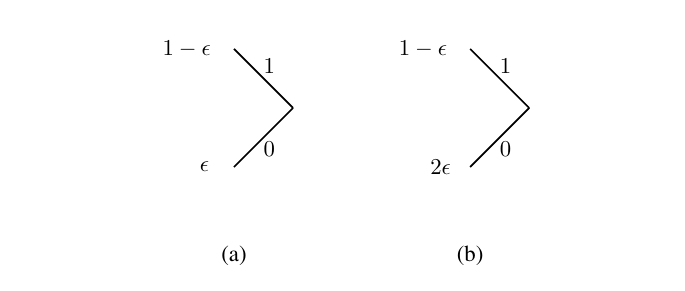}
\caption{Markov processes in Example \ref{ergodic_ex} with stationary probabilities (a) $\stn(1)=\stn(0)=\frac{1}{2}$  (b) $\stn'(1)=\frac{2}{3}, \stn'(0)=\frac{1}{3}$. Given a sample with size $n$ with $\epsilon \ll o(1/n)$, we cannot distinguish between these two models.}
\label{ergodic_ex_fig}
\end{center}
\end{figure}

\bExample Let $\A=\{ 0,1\}$ and $\T=\{ 0,1 \}$ with $q(1|1)=1-\epsilon$, and
$q(1|0)=\epsilon$. For $\epsilon>0$, this model represents a stationary ergodic Markov process with stationary
distributions $\stn(1)=\frac{1}{2}, \stn(0)=\frac{1}{2}$. Let $\T'=\{ 0,1 \}$ with $q'(1|1)=1-\epsilon,
q'(1|0)=2\epsilon$. Similarly, for $\epsilon>0$ this model represents a stationary ergodic Markov process with
stationary distributions $\stn'(1)=\frac{2}{3}, \stn'(0)=\frac{1}{3}$.

Suppose we have a length-$n$ sample and suppose $\epsilon\ll 1/n$.  If
we start from 1 (or 0), both models will yield a sequence of $n$ 1's
(or 0's) with high probability. Therefore, the length $n$ samples from
the two sources look identical. Hence no estimator could distinguish
between these two models with high probability if $\epsilon \ll
o(1/n)$, and therefore no estimator can obtain their stationary probabilities
either.
\label{ergodic_ex}\eExample

Finally there is, of course, no guarantee that the counts of short
strings are more amenable to interpretation than longer ones in a
long-memory, slow mixing process.
  
\bExample Let $\T=\{11,01,10,00\}$ with $q(1|11)=\epsilon$, $q(1|01)=\frac{1}{2}$, $q(1|10)=1-\epsilon$,
$q(1|00)=\epsilon$. If $\epsilon>0$, then $\ptq$ is a stationary ergodic binary Markov process. Let $\stn$ denote the
stationary distribution of this process. A simple computation shows that \ignore{
\begin{align*}
Q=\begin{bmatrix}
  \epsilon & 0 & 1-\epsilon & 0 \\
  \frac{1}{2} & 0 & \frac{1}{2} & 0 \\
  0 & 1-\epsilon & 0 & \epsilon \\
  0 & \epsilon & 0 & 1-\epsilon
 \end{bmatrix}
\end{align*}
and} $\stn(11)=\frac{1}{7-6\epsilon}$,
$\stn(01)=\frac{2-2\epsilon}{7-6\epsilon}$,
$\stn(10)=\frac{2-2\epsilon}{7-6\epsilon}$ and
$\stn(00)=\frac{2-2\epsilon}{7-6\epsilon}$, and
$\stn(1)=\frac{1}{7-6\epsilon}+\frac{2-2\epsilon}{7-6\epsilon}=\frac{3-2\epsilon}{7-6\epsilon}$
and
$\stn(0)=\frac{2-2\epsilon}{7-6\epsilon}+\frac{2-2\epsilon}{7-6\epsilon}=\frac{4-4\epsilon}{7-6\epsilon}$.

\begin{figure}
\centering
\includegraphics[trim=5cm 12cm 5cm 11cm, clip,scale=.7]{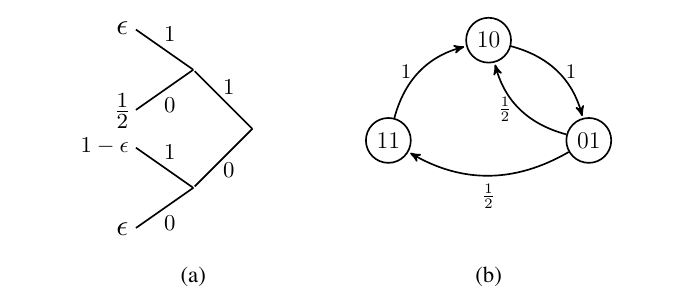}
\caption{(a) Markov in Example \ref{dependency1}, (b) Same process when $\epsilon=0$ .}
\label{figure:ergodic_mode}
\end{figure}

Suppose we have a length $n$ sample. If $\epsilon \ll \frac{1}{n}$,
then $\stn(1)\thickapprox\frac{3}{7}$ and
$\stn(0)\thickapprox\frac{4}{7}$. If the initial state belongs to
$\{11,01,10\}$, the state $00$ will not be visited with high
probability in $n$ samples, and it can be seen that the counts of $1$
or $0$ will not be near the stationary probabilities $\stn(1)$ or
$\stn(0)$. For this sample size, the process effectively acts like the
irreducible, aperiodic Markov chain in
Fig. \ref{figure:ergodic_mode}. (b) which can be shown to be fast
mixing. Therefore the counts of 01, 10 and 11 approach the stationary
probabilities of the chain in Fig. \ref{figure:ergodic_mode}. (b), namely
$\frac{\stn(01)}{\stn(1)+\stn(10)}$,
$\frac{\stn(10)}{\stn(1)+\stn(10)}$, and
$\frac{\stn(11)}{\stn(1)+\stn(10)}$, much quicker than the counts of
1 and 0 will approach $\stn(1)$ or
$\stn(0)$. Indeed, this observation guides our search for results in
Section \ref{Dev_bound_Naive}.
\label{dependency1}
\eExample 

\section{Summary of results}
\label{s:rsl}
\ignore{Motivated by the operation and estimation in backplane channels, we consider the following abstraction.} We observe a length
$n$ sample from a stationary, ergodic, $\cA-$ary Markov source $\ptq$, where both $\cT$ and $q(\cT)$ are unknown. Using
this sample, we want (i) to approximate as best as possible, the parameter set $q\Paren\cT$ (ii) the stationary
probabilities $\stn(\s)$ of strings $\s \in\cT$, and (iii) to estimate or at least obtain heuristics of the entropy rate
of the process. 

Two distinct problems complicate estimation of $q\Paren{\cT}$ and the
stationary probabilities. First is the issue that the memory may be
too long to handle---in fact, if the source has long enough memory it
may not be possible, with $n$ samples, to distinguish the source even
from a memoryless source (Example \ref{dependency}). Second, even if
the source has only one bit of memory, it may be arbitrarily slow
mixing (Example~\ref{ergodic_ex}). No matter what $n$ is, there will
be sources against which our estimates perform very poorly.

Given $Y^0_{-\infty}$, we get the sample sequence $Y_1^n$ from the
(unknown) model, namely $Y_i$ is generated with probabilities
$q(Y_i|\ctxt(Y^{i-1}_{-\infty}))$.  Since we do not know $\cT$, a
natural way to proceed is to estimate conditional probabilities of
form $\ptq(Y_1|\U)$, where $\U\in\cA^\kn$ are strings of a given
length $\kn$.  Thus, we obtain a potentially coarser model with
states $\Ttilde=\cA^\kn$ for some known $\kn$.  With the benefit of
hindsight, we take $\kn=\cO(\log n)$\footnote{A function
  $f_n=\cO(g_n)$ if $\exists n_0 \in \mathbb{N}$ and $\exists M>0$
  such that $f_n\leq M g_n$ for $n\geq n_0$.} and write
$\kn=\alpha_n\log n$ for some function $\alpha_n=\cO(1)$. This scaling
of $\kn$ also reflects well known conditions for consistency of
estimation of Markov processes in~\cite{CT06}.  \ignore{This will be
  the high level approach we will take as well. The particular value
  of $\alpha_n$ will not be as important as the fact that we take it
  $\cO(1)$.}

For convenience, we rephrase the above problem by defining an
\textit{aggregation} of a Markov process in Section \ref{Sec4} at
depth $\kn$. The aggregation of the true model can be thought of a
coarse approximation of the true model---the aggregated model has
memory $\kn$ and (unknown) parameters
associated with $\U\in\cA^\kn$, $\tilde{q}(a|\U)$, set to $\ptq(a|\U)$.
We denote the aggregated model by $\ptildettildeq$ (with states $\Ttilde$ 
and parameters $\tilde{q}(\Ttilde)$).
In Proposition~\ref{UB_Ent_Rate}, we show that the
entropy rate corresponding to the aggregated model $\ptildettildeq$ is
an upper bound on the entropy rate of the true model $\ptq$.

The catch is that we do not get to see observations corresponding to
the aggregated model. We need to estimate the aggregated model
$\ptildettildeq$ using the observations from the true, underlying
model $\ptq$. Therefore the task is not the same as estimating a model
with memory $\kn$.

\ignore{\paragraph*{Entropy rate}
We first compare the entropy rates of the true model with that of the
aggregated model. This is of course, a purely hypothetical
comparison---it is not as if we get to observe the aggregated
model. 
While it may not always be possible to directly use
Proposition~\ref{UB_Ent_Rate} to estimate entropy rates of slow mixing
Markov models, we develop the notion of a \emph{partial entropy rate},
a useful heuristic that is perhaps the best possible with a finite
sample. In channel estimation problems of~\cite{APS13}, the analogs
are in the form of heuristics on information rates (with the
aggregated channel models providing a lower bound on the true
information rates). Moreover, Proposition~\ref{UB_Ent_Rate} also
motivates the estimation questions that form our main results.}

\paragraph*{Naive estimates} To obtain the parameters of the
aggregated process, $\tilde{q}(\Ttilde)$, suppose we use a
\emph{naive} estimator that proceeds as though the sample was in fact
generated from $\ptildettildeq$. Namely the naive estimator is based
on the premise that the subsequence of symbols in the sample following
any $\w\in\Ttilde$ is \iid.  Consider the following
illustration. \ignore{Superficially, }

Suppose the binary Markov sample is 1101010100. Let
$Y_{-\infty}^{0}=\cdots 00$. We want to estimate the aggregated
parameters at depth 2---the conditional probability that $a$ follows
given a two bit string. In particular, say we want the aggregated
parameters associated with string 10,
$\tilde{q}(a|10)=\ptq(a|10)$. The subsequence following 10 in the
sample is 1110. Then the naive estimate of $\tilde{q}(1|10)$, denoted
by $\hat{q}(1|10)$, is 3/4 and the naive estimate of
$\tilde{q}(0|10)$, denoted by $\hat{q}(0|10)$, is 1/4.

There are two ways the naive estimate may still work. (i) The counts
of 101 and 100 reflect their stationary probabilities, which
automatically means that the count of 10 represents its stationary
probability as well.  (ii) The subsequence 1110 that follows 10 is
\iid. Case (i) is not valid since we have not assumed that the source
has mixed. Case (ii) would only hold if the string $10\in\T$ as well
or if the sample was from the aggregated source at depth 2---neither
assumption is justified at this point. \emph{Therefore, in general
  there is no reason why the naive approach even makes sense.}
However, assuming dependencies die down as below, we show that certain
naive estimates still capture the conditional probabilities
accurately.

\paragraph*{Dependencies die down} However, it is reasonable given our motivation that the influence of prior symbols dies down as
we look further into the past---the incremental value of an additional
page in browser history diminishes when the amount of history we already
have access to increases. 

We formalize this notion in Section \ref{EST_DEV_B} with a function
$d(i)$ that controls how symbols $i$ locations apart can influence
each other, and require $\sum_{i\geq 1} d(i) <\infty$. Let $\cM_d$ be
the set of all models $\ptq$ that satisfy for all $\U\in\cA^*$ and all
$b,b' \in \A$ and for all $a\in \A$.
\begin{equation*}
\quad \bigg |\frac{\ptq(a|b\U)}{\ptq(a|b'\U)}-1 \bigg|\leq d(|\U|).
\end{equation*}
It can be easily shown that the mutual information between bits $i$
apart, conditioned on all bits between them, is upper bounded by $\log
(1+d(i))$.  Note that there is no bound on the memory of models in
$\cM_d$. Moreover, the function $d$ does not constrain mixing
properties of the processes as we show in Section~\ref{EST_DEV_B}.

\paragraph*{Conditional probabilities}
For sources $\ptq\in\cM_d$, we show how to obtain (from the data
sample) $\tilde{G}\subseteq \Ttilde$, a set of \emph{good
  states}\footnote{Strings in $\Ttilde$ may not be states of $\ptq$,
  but we abuse this notation for convenience.} or \emph{good}
length-$\kn$ strings (Definition \ref{Good_States}) from sample
sequence. These are strings that will be amenable to concentration
results, and hence the adjective ``good''. These results do not depend
on empirical counts of strings being near their stationary
probabilities, nor do they require that the subsequence following a string
$\w\in\Ttilde$ be \iid.

\noindent
\paragraph*{\bf{Main Result 1}} In Theorem \ref{thm:aggemptht}, we
show that with probability (under the underlying unknown $\ptq$) $\ge
1-\frac1{2^{|\A|^{\kn+1}\log n}}$ (conditioned on any past
$Y_{-\infty}^0$), for all states $\w\in\Ttilde$ simultaneously
\begin{equation}
  \Vert\tilde{q}(\cdot|\w) -\hat{q}(\cdot|\w)\Vert_{_1} \le 2\sqrt{\frac{(\ln 2) (|\A|^{\kn+1}\log n+n\delta_\kn)}{N_{\w}}}.
\label{eq:rsl:tp}
\end{equation}
Here, $\delta_{\kn}=\sum_{i \geq k_n} d(i)$, $N_{\w}$ is the number of
occurrences of string $\w$ in the sample and $\hat{q}(.|\w)$ is the
naive estimator of $\tilde{q}(.|\w)$ as described above. 
\ignore{The right side of \eqref{eq:rsl:tp} diminishes to $0$ for all $\w \in \tilde{G}$. }
Note that \eqref{eq:rsl:tp} automatically yields sharper
estimates for those strings $\w$ whose counts are larger. 

For example, if $d(i)=\gamma^i$ for some $0<\gamma<\frac{1}{2}$, then
$\delta_\kn=\frac{\gamma^\kn}{1-\gamma}$. By choosing $\kn=\gamma \log
n$, the accuracy of $\hat{q}(\cdot|\w)$ is $ \Theta\big(\sqrt{\frac{
    n^{1+\gamma \log \gamma}}{N_{\w}}}\big)$. In particular, for a
strings appearing at least $\frac{n}{\log n}$ times, the accuracy of
estimation in \eqref{eq:rsl:tp} is better than
$\Theta\big(\sqrt{n^{\gamma\log \gamma}\log
  n}\big)$. \ignore{footnote{A function $f_n=\Theta(g_n)$ if
    $f_n=\cO(g_n)$ and $g_n=\cO(f_n).$}}

The above estimation result is built on two facts: (i) dependencies
dying down and (ii) universal compression results on $\cM_d$ built on
the fact that length-$n$ sequences generated by Markov sources with
memory $\kn$ can be universally compressed if $k_n=\cO(\log n)$.

A related curiosity arises due to the fact that the above result does
not depend on empirical frequencies being close to stationary
probabilities. The result above is sometimes tight for strings $\w$
while being vacuous for their suffixes $\w'$. For example, it could be
that we estimate parameters associated with a string $\w$ of length
$\Theta(\log n)$ (say $\w$ is a string of ten 0's) but not those
associated with $\w'$ where $\w'\preceq \w$ (say $\w'$ is a string of
five 0's).  Finally, since the above result is what can be obtained without
any knowledge of mixing properties,~\eqref{eq:rsl:tp} could be
strengthened using other arguments in literature in cases where we may
know that empirical counts reflect stationary probabilities.

\paragraph*{Stationary probabilities of strings}
In general, stationary probabilities of strings can be a very
sensitive function of the transition probabilities. We now have the
approximate transition probabilities associated with strings in
$\Gtilde$. With this little bit of information we have gleaned from
the sample, can we even hope to say anything about stationary
probabilities of $\w\in\Gtilde$?  How then do we interpret the
empirical counts $N_\w$ of various strings $\w$? 

To answer this question, we calculate a parameter $\cplngg$
in~\eqref{eq:cplngg} which resembles the Dobrushin's ergodicity
coefficient of Markov processes, but which can be estimated well
using~\eqref{eq:rsl:tp}. Suppose $\{\delta_i\}_{i\geq 1}$ is summable
as well, and let $\Delta_j=\sum_{i\geq j} \delta_i$.

\noindent
\paragraph*{\bf{Main Result 2}}
In Theorem \ref{stat-estim} we show under a minor technical condition
that for all $t>0$, $Y^0_{-\infty}$ and $\w\in {\tilde G}$ the counts
of $\w$ in the sample, $N_{\w}$, concentrates though not necessarily
around $\stn(\w)$. We show that
\begin{equation}
\label{eq:rsl:sp}
\ptq (|N_{\w}-\tilde{n}\frac{\stn(\w)}{\stn(\tilde{G})} \,|\geq t | Y^0_{-\infty}) 
\leq 
2\exp 
\Paren{
-\frac{(t-\cB)^2}{2\tilde{n}\cB^2}},
\end{equation}
where
$\cB\approx4\max\sets{\ell_n,\kn}/[\cplngg^\kn(1-\Delta_{\kn})]$. Here,
$\ell_n$ is the smallest integer such that $\Delta_{\ell_n}\leq
\frac{1}{n}$, $\tilde{n}$ is the total count of good states in the
sample and $\stn$ denotes the stationary distribution of $\ptq$.  Note
once again that $\ptq$ is the probability law under the underlying
unknown model in $\cM_d$.  The above estimation
result~\eqref{eq:rsl:sp} uses a coupling argument~\cite{A83} to bound
martingale differences of a natural Doob martingale construction in
Section~\ref{Dev_bound_Naive}.

When dependencies $d(i)$ decay exponentially, we will have
$\ell_n=\Theta(\log n)$.  Suppose $\ntilde=\cO(n)$ and we want our
confidence to approach 1 polynomially in $n$. If $\cplngg$ specified
in Section~\ref{Dev_bound_Naive} is $\Theta(1)$ and sufficiently
large, it implies reasonable mixing within the good states.  In this
case, there will be $0<\beta<\frac{1}{2}$ such that for all
$\w\in\Gtilde$, $N_w/\tilde{n}$ is within $\frac{(\log
  n)^2}{n^{1/2-\beta}}$ from the ratio $\stn(\w)/\stn(\Gtilde)$ with
the required confidence. In case $\cplngg$ turns out to be too small
to yield good deviation bounds, one can either shrink $\Gtilde$ to
include a subset of states that mix well, or move to smaller values
for $\kn$ (but still scaling as $\Theta(\log n)$).

To summarize, note that $\kn$, $\Delta_\kn$ and $\ell_n$ (the later
two related to how fast dependencies die down) are known
$a-priori$. But $\tilde{n}$ is a random variable found from the
sample. So is $\cplngg$, but one that can be well estimated from the
sample using~\eqref{eq:rsl:tp}.  \ignore{(simultaneously well for all
  contexts in $\Gtilde$ with confidence $\ge
  1-\frac1{2^{|\A|^{\kn+1}\log n}}$) It is thus perfectly possible
that for some samples, we may be able to say little about the counts
$N_\w$, but in other samples we could interpret it well.}
The result~\eqref{eq:rsl:sp} is a natural deviation bound where the
confidence (right side of~\eqref{eq:rsl:sp}) is a random variable
generated from the model $\ptq$, but one that can be well estimated
from the sample (the confidence in~\eqref{eq:rsl:sp} is a decreasing
function of $\cplngg$). To use~\eqref{eq:rsl:sp} when confronted with
a sample, we lower bound $\cplngg$ by $\bar{\eta}_{{}_{\tilde{G}}}$
with confidence $\ge 1-\frac1{2^{|\A|^{\kn+1}\log n}}$
using~\eqref{eq:rsl:tp} to conservatively obtain a further upper bound
on the left side of~\eqref{eq:rsl:sp}.

\ignore{With the effect that
the model dependent right side is replaced by another upper
bound---potentially worse, but entirely data dependent (with a new
reduced confidence obtained by a union bound on~\eqref{eq:rsl:tp}
and~\eqref{eq:rsl:sp}).}

\bRemark All logarithms are base 2. We use bold font $\w$ or $\s$ for
strings. Typically $\s$ is a generic state or context of a process,
while $\w$ is used for a ``good'' state as a mnemonic. A subscript
$\s$ usually refers to an instance of the Markov process whose past
corresponds to $\s$---for example, $N_\s$ (for the count of $\s$ in
the sample---the number of times the sample had $\s$ as its immediate
past).  \eRemark

\section{Background}
\subsection{Context tree weighting} 
\label{s:ctxtree}
Context tree weighting is a universal data compression algorithm for
Markov sources~\cite{WST95,TSW93} that captures several insights on
how Markov processes behave in non-asymptotic regimes. Let $y_1^n$
be sequence of symbols from an alphabet $\A$. Let $\hat{\T}=\A^K$ for
some positive integer $K$. For all $\s \in \hat{\T}$ and $a \in \A$,
let $\nsa$ be the number of $a$'s which appear exactly after the string $\s$ in $y_1^n$. The
depth-$K$ context tree weighting constructs a distribution $\pctxtree$
satisfying~\footnote{Note that the bound holds for $n \geq 2$.}
\begin{equation*}
\pctxtree (y_1^n|y_{-K+1}^0) 
\geq 
2^{-|\A|^{K+1} \log n} 
\prod_{\s \in \hat{\T}} 
\prod_{a\in \A} 
{ \bigg( 
\frac{\nsa}{\sum_{a \in \A}{\nsa}}\bigg) ^{\nsa} }.
\end{equation*}
The above inequality is not the strongest, but its form is convenient for use.
Note that no Markov source with memory $K$ could have given a higher probability to $y_1^n$ than
\[ 
\prod_{\s \in \hat{\T}} 
\prod_{a\in \A} 
{ \bigg( \frac{\nsa}{\sum_{a \in \A}{\nsa}}\bigg) ^{\nsa} }.\]
So, if $|\A|^K\log n=o(n)$, then $\pctxtree$ underestimates any memory-$K$ Markov probability by only a subexponential
factor. Therefore, $K=\cO(\log n)$ is going to be the case of particular interest.

\subsection{Coupling for Markov processes}
\label{s:b:cpl}
We adopt the coupling~\cite{A83} technique in
Section~\ref{Dev_bound_Naive} to estimate stationary
probabilities. Coupling is an elegant approach to interpret counts of
certain strings in a sample.  Let $\ptq$ be our Markov source
generating sequences from an alphabet $\tilde{\cA}$.

A coupling $\coplg$ for $\ptq$ is a special kind of joint probability
distribution on the sequences $\{Y_m,\bar{Y}_m\}_{m\ge1}$ where
$Y_m\in\tilde{\cA}$ and $\bar{Y}_m\in \tilde{\cA}$. $\coplg$
has to satisfy the following property: individually taken, the sequences
$\{Y_m\}$ and $\{\bar{Y}_m\}$ have to be faithful evolutions of
$\ptq$. Specifically, for $m\ge0$, we want
\begin{align}
\label{eq:cplng}
\coplg(Y_{m+1}| Y_{-\infty}^m,\bar{Y}_{-\infty}^m)
&=
\ptq(Y_{m+1}| Y_{-\infty}^m) \nonumber \\
&=
\ptq(Y_{m+1}| \ctxt(Y_{-\infty}^m)),
\end{align}
and similarly for $\{\bar{Y}_m\}$. 

In the context of this paper, we think of $\sets{Y_m}$ and $\sets{\bar{Y}_m}$ as copies of
$\ptq$ that were started with two different states $\s, \s' \in \T$
respectively, but the chains evolve jointly as $\coplg$ instead
of independently. For any $r$ and $\w \in {\tilde{\cA}}^r$, $N_\w$
(respectively $\bar{N}_\w$) is the number of times $\w$ forms the context of
a symbol in a length-$n$ time frame, $\sets{Y_i}_{i=1}^n$ given $Y^0_{-\infty}$
(respectively $\sets{\bar{Y}_i}_{i=1}^n$ given ${\bar Y}^0_{-\infty}$). Then, for any $\coplg$~\footnote{\hrule For any event $E$, the indicator function is defined as \[\indctr(E)= \begin{cases} 1 & \textrm{ if $E$ holds,} \\
      0 & \textrm{otherwise.} \\
   \end{cases}.\]},
\begin{align*}
\big \vert 
&\E_{\ptq}[ N_\w| Y^0_{-\infty}] 
- 
\E_{\ptq}[\bar{N}_\w| {\bar Y}^0_{-\infty}] \big \vert \\
&=
\bigg \vert 
\sum_{i=1}^{n} 
\E_\coplg
\big[
\indctr
\big(
\ctxs{\tilde{\cA}^r}{(Y^i_{-\infty})}=\w\big) 
-
\indctr\big(\ctxs{\tilde{\cA}^r}{(\bar{Y}^i_{-\infty})}=\w\big)\big]
\bigg \vert\\
&\leq  
\sum_{i=1}^{n} 
\bigg \vert 
\E_\coplg\big[
\indctr\big(\ctxs{\tilde{\cA}^r}{(Y^i_{-\infty})}=\w\big) 
-
\indctr\big(\ctxs{\tilde{\cA}^r}{(\bar{Y}^i_{-\infty})}=\w\big)
\big] 
\bigg \vert \\
&\leq  
\sum_{i=1}^{n}  
\coplg\big(
\ctxs{\tilde{\cA}^r}{(Y^i_{-\infty})} 
\neq 
\ctxs{\tilde{\cA}^r}{(\bar{Y}^i_{-\infty})}
\big),
\end{align*}

where the first equality follows from~\eqref{eq:cplng}. 

The art of a coupling argument stems from the fact that $\coplg$ is
completely arbitrary apart from having to satisfy~\eqref{eq:cplng}. If
we can find \emph{any} $\coplg$ such that the chains \emph{coalesce},
namely $\coplg\big(\ctxs{\tilde{\cA}^r}{(Y^i_{-\infty})} \neq
\ctxs{\tilde{\cA}^r}{(\bar{Y}^i_{-\infty})}\big)$ becomes small as $i$
increases, then we know that $\E_{\ptq}[ N_\w| Y^0_{-\infty}] $
cannot differ too much from $\E_{\ptq}[\bar{N}_\w| {\bar Y}^0_{-\infty}]$. 

Now if, in addition, such a coalescence holds no matter what 
$ {\bar Y}^0_{-\infty}$ is, we could then pick $ {\bar Y}^0_{-|\cT|}$
according to the stationary distribution of $\ptq$. Then $\bar{N}_\w$ 
would be close to the stationary count of $\w$,
and from the coupling argument above, so is $N_\w$.
For tutorials, see
\eg~\cite{FG,LPW09,G00}.

\section{Model Aggregation}
\label{Sec4}
\subsection{Entropy Rate}
The entropy rate of a stationary Markov process $\ptq$, denoted by $\entrT$, is defined as~\cite{CT}
\begin{equation*}
\entrT= -\sum_{\s \in \T} \stn(\s) \sum_{a \in \A} q(a|\s) \log q(a|\s) \ed \sum_{\s \in \T} \stn(\s) \hs.
\end{equation*}
\subsection{Aggregations}
\begin{figure}[!t]
\begin{center}
\includegraphics[trim=5cm 12cm 5cm 11cm, clip,scale=.7]{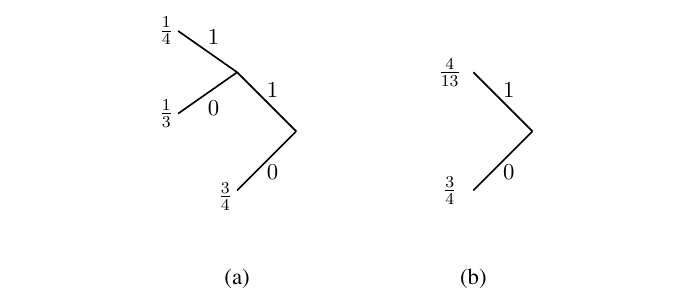}
\caption{(a) Markov process in Example \ref{aggexample}, (b) Aggregated model at depth 1. From Observation 1, the 
model on the left can be reparameterized to be a complete tree at any depth $\ge 2$. We can hence ask for its aggregation at any
depth. Aggregations of the above model on the left at depths $\ge2$ will hence be the model itself.}
\label{aggfig}
\end{center}
\end{figure}

Since the memory is unknown a-priori, a natural approach,
known to be consistent, is to use a potentially
coarser model with depth $k_n$. Here, $k_n$ increases logarithmically 
with the sample size $n$, and reflects~\cite{CT06} well known results on consistent
estimation of Markov processes.
We show that coarser models formed by properly aggregating states of the
original context tree model are useful in upper bounding entropy rates of
the true process.
\bDefinition
Suppose $\Ttilde=\A^k$ for some positive integer $k$. 
The \textit{aggregation} of $\ptq$ at level $k$, denoted by $\ptildettildeq$, 
is a stationary Markov process with state transition probabilities given by
\begin{equation}
\tilde{q}(a|\s)
=
\ptq(a|\s)
=
\frac{\sum_{\V \in \T_\s} \stn(\V)q(a|\V)} {\sum_{\V' \in \T_\s} \stn(\V')},
\label{trans_agg}
\end{equation}
for all $\s \in \Ttilde$ and $a \in \A$, where $\stn$ is the stationary distribution associated with $\ptq$.
\label{Aggregation}
Using Observation~\ref{obs:indp}, wolog, no matter what $\Ttilde$ is,
we will assume $\ptq$ has states $\cT$ such that $\Ttilde\preceq\cT$.
\eDefinition
\bExample
\label{aggexample}
This example illustrates the computations in Definition above. Let $\ptq$ be a binary Markov process with $\T=\{11, 01, 0\}$ and $q(1|11)=\frac{1}{4}, q(1|01)=\frac{1}{3}, q(1|0)=\frac{3}{4}$.
For this model, we have $\stn(11)=\frac{4}{25}, \stn(01)=\frac{9}{25}$ and $\stn(0)=\frac{12}{25}$.
Fig. \ref{aggfig}. (b) shows an aggregated process $\ptildettildeq$ with $\Ttilde=\{1, 0\}$. 
Notice that $\tilde{q}(1|1)=\big(\frac{4}{25}\frac{1}{4}+\frac{9}{25}\frac{1}{3}\big)/(\frac{4}{25}+\frac{9}{25})=\frac{4}{13}$ and $\tilde{q}(1|0)=\frac{3}{4}$.
\eExample

\bLemma
Let $\ptq$ be a stationary Markov process with stationary distribution $\stn$. If $\ptildettildeq$ aggregates $\ptq$ then
it has a unique stationary distribution $\tilde{\stn}$ and for every $\w \in \tilde{\mathcal{T}}$
$$\tilde{\stn}(\w)=\sum_{\V \in \mathcal{T}_{\w}} \stn(\V).$$
\ignore{Moreover, for all $a_1^k \in \mathcal{A}^k$ such that $a_1^k$ is an internal node of $\Ttilde$ we have $\tilde{\stn}(a_1^k)=\stn(a_1^k)$.} 
\label{aggregated model}
\Proof See Appendix~\ref{s:agg}.
\eLemma 

\subsection{Upper Bound}
Suppose $\ptildettildeq$ aggregates $\ptq$,
and let the entropy rate of the aggregated process be 
$\entrTt$. Then,
\bProposition
$\entrT \leq \entrTt$.
\label{UB_Ent_Rate}
\Proof See Appendix~\ref{s:upper}.
\eProposition

\bRemark In this paper, we are particularly concerned with the slow mixing regime.  As our results will show, in general
it is not possible to obtain a simple upper bound on the entropy rate using the data (given a particular starting state)
and taking recourse to the Proposition above.  Instead, we introduce the \textit{partial} entropy rate that can be reliably
obtained from the data
\[
\mathcal{H}_{{ }_{\tilde{G}}}=\sum_{ \substack{\w \in {\tilde G}}} \frac{\stn({\w})}{\stn(\tilde{G})}\hw,
\]
where ${\tilde G}\subseteq\tilde{\cT}$ will be a set of \emph{good} states
that we show how to identify.
Recall that $\stn(\tilde{G})$ may be arbitrarily small in a finite sample, hence we must expect to compute $\mathcal{H}_{{ }_{\tilde{G}}}$ and not $\stn(\tilde{G})\mathcal{H}_{{ }_{\tilde{G}}}$.
The partial entropy rate is not necessarily an upper bound, but in
slow mixing cases it is sometimes the best heuristic possible. We
systematically handle the entropy rates of slow mixing processes 
using the estimation results below in different paper.
\eRemark 
\ignore{
Notwithstanding the previous remark, we will focus on estimating the aggregated
parameters $\tilde{q}(\Ttilde)$, where $\Ttilde=\cA^\kn$ has depth $\kn$ where $k_n$ grows logarithmically as $\alpha_n \log n$ for some $\alpha_n=\cO(1)$.
\bRemark
The aggregated model defined in this section is purely to facilitate our
analysis. We never get to observe the aggregated model, though we will 
pose estimation problems in terms of its parameters. 
We only can observe samples from the underlying (original as opposed to aggregated) model. From these observations
we must figure out $q(\T)$ if at all possible. In general, there 
is no real reason why it should even be possible. But we will be able 
to handle this task because of~\eqref{eq:condition}.
\eRemark }

\section{Dependencies in the Model}
\label{EST_DEV_B}
As noted before in Example \ref{dependency}, if the
dependencies could be arbitrary in a Markov process, we will not
estimate the model accurately no matter how large the sample
is. 
Keeping in mind Observation~\ref{obs:indp}, we formalize dependencies dying down by means of a function
$d:\mathbb{Z}^{+}\to \mathbb{R}^+$ with $\sum_{i=1}^{\infty} d(i) <\infty$.

Let $\cM_d$ be the set of all models $\ptq$ that satisfy for all $\U\in\cA^*$ and all $b,b'
\in \A$ and for all $a\in \A$. 
\begin{equation}
\label{eq:condition}
\quad \bigg |\frac{\ptq(a|b\U)}{\ptq(a|b'\U)}-1 \bigg|\leq d(|\U|).
\end{equation}
Note that our collection $\cM_d$ has bounded memory iff there exists a
finite $K$ such that $d(i)=0$ for all $i>K$. 
\bRemark We emphasize that the restriction $\{d(i) \}_{i\geq 1}$ does not preclude slow mixing processes in
$\cM_d$---we clarify this point in following example.  
The restriction $\{ d(i)\}_{i\geq 1}$ we use is related to
the notion of ``continuity rate'' of stochastic processes used
in~\cite{DGG06,GF08,CGL08,CT10}. 
\eRemark

\bExample Let $\ptq$ be a context tree model with $\T=\{0,1\}$ and transition probabilities $q(1|1)=1-\epsilon, \,q(1|0)=\epsilon$ as in Fig. \ref{ergodic_ex_fig}. (a). A simple calculation shows that the stationary probability of being at state 0 or 1 is $\frac{1}{2}$.
Note that even with very strong restriction on $d$, namely
$d(i)=0$ for $i\ge1$, $(\T,q(\T))$ belongs to $\cM_d$ regardless of the value of $\epsilon$.

While we do not need the notions of $\phi$-mixing and $\beta$-mixing coefficients for stationary stochastic processes \cite{Yu94} in the rest of the paper, we compute them for this example as an illustration that $\{d(i)\}_{i\geq 1}$ are unrelated to mixing. 

Recall that the $j'$th $\phi$-mixing coefficient  
\begin{align*}
\phi_j 
&\ge
\max \limits_{Y_0 \in \sets{0,1}} 
\max \limits_{Y_j \in \sets{0,1}} 
\big |\ptq(Y_j|Y_0)-\ptq(Y_j)\big| \\
&\geq 
\big |\ptq(Y_j=1|Y_0=1)-\ptq(Y_j=1) \big | \\
&\geq
\ptq(Y_j=1,Y_{j-1}=1,\cdots, Y_1=1|Y_0=1)\\
&-\ptq(Y_j=1)\\
&=
(1-\epsilon)^j-\frac{1}{2}.
\end{align*}
Similarly, the $j'$th $\beta-$mixing coefficient
\begin{align*}
\beta_j&\ge \E_{Y_0} 
\bigg[ 
\max \limits_{Y_j \in \sets{0,1}} 
|\ptq(Y_j|Y_0)-\ptq(Y_j)| 
\bigg]\\
&=
(1-\epsilon)^j-\frac{1}{2}.
\end{align*}
No matter what $\epsilon>0$ is, $\ptq\in \cM_d$ even under stringent restriction $d(i)=0$ for $i\geq 1$. But,
any $(\beta,\phi)$-mixing coefficient can be made arbitrarily as close to $\frac{1}{2}$ as possible by picking
$\epsilon$ small enough. Therefore, the condition \eqref{eq:condition} we impose does not preclude slow mixing.  \eExample

As mentioned in the last section, we will focus on set of the
aggregated parameters at depth $\kn$, $\tilde{q}(\cA^\kn)$ where
$k_n=\alpha_n \log n$. If $k_n$ is large enough, these aggregated
parameters start to reflect the underlying parameters $q(\T)$. Indeed,
by using an elementary argument in Section \ref{est_ch_par} we will
show that both the underlying and aggregated parameters will then be
close to the empirically observed values for states that occur
frequently enough---even though the sample we use comes from the true
model $\ptq$ 
instead of the aggregated model
$\ptildettildeq$.
\bRemark
In the context of information aggregation on the Internet, suppose we have a browser
history of the $i$ prior webpages (or more appropriately domains) visited. What
is the incremental value of the $(i+1)'$th document in the past? It
is reasonable to assume that as $i$ increases, the incremental value of another
past document diminishes. It is easy to see using elementary arguments that
in a Markov process $\{Y_i\}$, $i\in\integers$ within the class $\cM_d$ that
\[
I(Y_0, Y_{i+1}|Y_1^{i})\le \log (1+d(i)),
\]
namely the function $d$ controls the incremental value of one additional data
point in the history. It is interesting to note that not all Markov processes
even in practical settings need to satisfy this observation (in fact problems in
DNA folding specifically cannot make this assumption), but that this assumption
is sufficient for statistical estimation problems to be well posed.  \eRemark

\bProposition Let $\{d(i)\}_{i\geq1}$ be a sequence of real numbers such that there exists some $n_0 \in \mathbb{N}$
for which, $0\leq d(i)\leq 1$ for all $i\ge n_0$.  Then, $\forall j\ge n_0$, we have 
\[
1-\sum_{i\ge j} d(i) \le \prod_{i\ge j}(1-d(i)) \le \frac1{\prod_{i\ge j} (1+d(i))}. 
\]
\label{Bonferroni}
\Proof See Appendix~\ref{s:f}.
\eProposition
\bProposition
Let $\ptq\in \cM_d$. Suppose $\ptildettildeq$ aggregates $\ptq$ with $\Ttilde=\A^{k_n}$. If $\sum_{i\geq k_n} d(i)\leq 1$, then for all $\w\in \Ttilde$ and $a\in \A$
\[
\bigg (1-\sum_{i\ge k_n} d(i) \bigg ) \max_{\s\in \cT_\w} q(a|\s)
\le 
\tilde{q}(a|\w)
\le
\frac{\min_{\s\in \cT_\w} q(a|\s)}
{\Paren{1-\sum_{i\ge k_n} d(i)}}.
\]
\Proof Recall that $\cT_\w$ are nodes of $\cT$ that are descendants of $\w\in\Ttilde$. See Appendix~\ref{s:die}.
\label{die_out}
\eProposition
\section{Naive Estimators}
\label{naive_est}
Even in the slow mixing case, we want to see if any estimator can be accurate at least partially. In particular, we consider the naive estimator that operates on the assumption that samples are from the aggregated model $\ptildettildeq$. There is no reason that the naive estimates should reflect the parameters associated with the true model $\ptq$.
Even in the slow mixing case, we want to see if any estimator can be
accurate at least partially. In particular, we consider the naive
estimator that operates on the assumption that samples are from the
aggregated model $\ptildettildeq$.  Without our assumption on dependencies
falling off, there is no \emph{a-priori} reason that the naive
estimates should reflect the parameters associated with the true model
$\ptq$.

\bDefinition
Given a sample sequence $Y_1^n$, let $\tilde{\T}=\A^{k_n}$ with $k_n=\alpha_n \log n$ for some function $\alpha_n=\cO(1)$. For $\s \in \tilde{\T}$ let $\Y_\s$ be the sequence of symbols that follows the string $\s$. Hence, the length of $\Y_{\s}$ is 
\[
N_{\s}=\sum_{i=1}^{n} \indctr \{\ctxtt (Y_{-\infty}^{i-1})=\s\}.
\]
Therefore the number of $a's$ in $\Y_{\s}$ is 
\[
\nsa=\sum_{i=1}^{n} \indctr \{\ctxtt (Y_{-\infty}^{i})=\s a\}.
\]
Observer that $N_{\s}=\sum_{a \in \A} \nsa$. The naive estimate of the aggregated parameters is
\[
\hat{q}(a|\s)=\frac{\nsa }{N_{\s}}.\eqed
\]
\label{impr_theta}
\eDefinitionp

\bRemark
Note that $\Y_\s$ is $\iid$ only if $\s \in \T$, the set of states for the true model. In general, since we do not necessarily know if any of $\nsa$ reflect the stationary frequencies, there is no obvious reason why $\hat{q}(a|\s)$ shall reflect $\tilde{q}(a|\s)$.
\eRemark \\

Since the process could be slow mixing, not all parameters are going to be accurate. Rather, there will be a set of good states in which we can do estimation properly.

Let $\delta_{j} = \sum_{i\ge j} d(i)$. Note that $\delta_j\to 0$ as $j\to\infty$ and 
that $-\delta_j\log\delta_j\to 0$ as $\delta_j\to 0$.
\bDefinition
\label{Good_States}
Given a sample sequence with size $n$ from $\ptq$, let  
\[
\tilde{G}=\big \{ \w \in \Ttilde:  \,N_{\w}\geq \max\sets{ n\delta_{\kn}\log\frac{1}{\delta_\kn}, |\A|^{\kn+1}\log ^2 n} \, \big\}.
\]
Note that the set $\Gtilde$ is obtained only using the sample and the
known function $d$. We call the set $\Gtilde$ ``good'' in an anticipatory fashion because we are
going to prove concentration results for estimates attached to these
states. 

Secondly, since we include arbitrary slow mixing sources, there is no way 
to estimate all conditional probabilities with a sample of length $n$. Therefore,
while the definition of $\Gtilde$ above may not be the tightest, some notion
of good states is unavoidable.
\eDefinition

\bRemark 
Throughout this paper, we assume that we start with some past
$Y^0_{-\infty}$, and we see $n$ samples $Y_1^n$ from $\ptq$. All confidence probabilities are conditional probabilities on
$Y_1^n$ obtained from underlying unknown model $\ptq$, given
past symbols $Y^0_{-\infty}$.  The results in Sections \ref{est_ch_par},~\ref{Dev_bound_Naive} hold for all
$Y^0_{-\infty}$ (not just with probability 1). In addition, if the history is not available, we can consider the first $\kn$ sample as the history and compute the empirical counts from the $n-\kn$ remaining samples. In other words, we can consider our length-$n$ sample as if we observe $Y_{-\kn+1}^{n-\kn}$ where $Y_{-\kn+1}^{0}$ can be thought as the needed history in computing the naive estimators. Since $\kn=\cO(\log n)$ is negligible compared to $n$, in the rest we skip this complication in computing empirical counts for the sake of simplification.
\eRemark
\section{Estimate of Transition Probabilities}
\label{est_ch_par}
Let $\Ttilde=\A^{k_n}$. Using samples from $\ptq$, we consider the
estimation of parameters $\tilde{q}(\tilde{\T})$ of the aggregated
model at depth $\kn$, and derive deviation bounds on the estimates in Theorem
\ref{thm:aggemptht}. However, before going to the proof of theorem, we
want to make the following important remark.

\bRemark Observe that because we do not assume the source has mixed,
the theorem below does not imply that the parameters are accurate for
contexts shorter than $\kn$. We may therefore be able handle longer
states' parameters (say a sequence of ten 0s), without being able to
infer those attached to their suffixes (say a sequence of five 0s).

This is perhaps counterintuitive at first glance. To see why this
could happen, note that the result below shows what can be obtained
without using anything about the mixing properties of the
source---namely, it does not rely on empirical frequencies of various
strings being close to their stationary probabilities. Therefore,
results on longer strings do not automatically translate to results on
shorter ones. Secondly, longer strings have attached conditional
probabilities closer to the true conditional probabilities with
which the source generates the data---therefore, there is less bias
to counter.

This is contrary to most prior work which obtain bounds on transition
probabilities subsequent to concentration of empirical counts of
strings around their stationary probabilities. Additional information
about the mixing of the source would further strengthen the following
results.
\eRemark

\bTheorem
\label{thm:aggemptht}
Let $Y_1^n$ be generated by an unknown model $\ptq \in\cM_d$. Let $k_n=\alpha_n \log n $.
Given any $Y^0_{-\infty}$, with probability under $\ptq$ 
$\ge 1-\frac1{2^{|\A|^{\kn+1}\log n}}$, 
for all $\w\in \Ttilde$ simultaneously
\[
D\big(\hat{q}(\cdot|\w) \Vert \tilde{q}(\cdot|\w) \big) \le \frac{2 (|\A|^{\kn+1}\log n+n\delta_\kn)}{N_{\w}}.
\]
\Proof As before, let $\delta_{\kn}=\sum_{i\ge \kn} d(i)$ and let $n$ be large enough that $\delta_{\kn}\leq \frac{1}{2}$.
Note that Proposition \ref{die_out} implies that for all sequences $y_1^n\in\cA^n$ and all $Y^0_{-\infty}$ 
\begin{align*}
\ptq(y_1^n|Y_{-\infty}^0)&\le \frac{1}{(1-\delta_{\kn})^n} \prod_{\w \in \Ttilde} \prod_{a \in \A} \tilde{q}(a|\w)^{\nwa}\\
&\le 
4^{n \delta_\kn}
\prod_{\w \in \Ttilde} \prod_{a \in \A} \tilde{q}(a|\w)^{\nwa},
\end{align*}
where $\nwa$ was defined in Definition~\ref{impr_theta} and the second inequality is because $(\frac{1}{1-t})^n \leq
4^{nt}$ whenever $0\leq t \leq \frac{1}{2}$. Now, let $B_n$ be the set of all sequences that satisfy
\begin{align*}
4^{n\delta_\kn} \prod_{\w \in \Ttilde} \prod_{a \in \A} \tilde{q}(a|\w)^{\nwa}
\le 
\frac{\prod_{\w \in \tilde{\T}} \prod_{a \in \A} \hat{q}(a|\w)^{\nwa}}{2^{2 |\A |^{\kn+1}\log n}}.
\end{align*}
Using a depth-$\kn$ context
tree weighting algorithm~\cite{TSW93} we obtain a distribution
$\pctxtree$ satisfying\footnote{While we use the context tree weighting algorithm, any worst case optimal universal compression algorithm would do for this theorem to follow.}
\[
\pctxtree(y_1^n|Y_{-\infty}^0) \ge \frac{\prod_{\w \in \Ttilde} \prod_{a \in \A} \hat{q}(a|\w)^{\nwa}}{2^{|\A|^{\kn+1}\log n}}.
\]
Now, for all sequences $y_1^n \in B_n$, we have
\begin{align*}
\pctxtree(y_1^n|Y_{-\infty}^0) &\ge
\frac
{\prod_{\w \in \Ttilde} \prod_{a \in \A} \hat{q}(a|\w)^{\nwa}}
{2^{|\A|^{\kn+1}\log n}} \\
& \ge \frac
{4^{n\delta_\kn}
\prod_{\w \in \Ttilde} \prod_{a \in \A} \tilde{q}(a|\w)^{\nwa}
2^{2|\A|^{\kn+1}\log n}}
{2^{|\A|^{\kn+1}\log n}} \\
&\ge
{\ptq(y_1^n|Y_{-\infty}^0)2^{|\A|^{\kn+1}\log n}}. 
\end{align*}
Thus, $B_n$ is the set of sequences $y_1^n$ such that $\pctxtree$ assigns a much higher probability than $\ptq$. Such a set $B_n$ can not have high probability under $\ptq$.
\begin{align*}
\ptq&(B_n)=\\
&\ptq\bigg\{ 
y_1^n: 
\pctxtree(y_1^n|Y_{-\infty}^0) 
\ge 
\ptq(y_1^n|Y_{-\infty}^0){2^{|\A|^{\kn+1}\log n}}
\bigg\}\\
&\leq \sum_{y_1^n \in B_n} \pctxtree(y_1^n|Y_{-\infty}^0) {2^{-|\A|^{\kn+1}\log n}}\\
&\le  {2^{-|\A|^{\kn+1}\log n}}.
\end{align*}
Therefore, with probability $\ge 1-2^{-|\A|^{\kn+1}\log n}$, (no matter $Y_{-\infty}^0$) we have
\[
\prod_{\w \in \Ttilde} \prod_{a \in \A} \tilde{q}(a|\w)^{\nwa}
\ge
\frac{
\prod_{\w \in \Ttilde} \prod_{a \in \A} \hat{q}(a|\w)^{\nwa}}
{2^{2|\A|^{\kn+1}\log n} \, 4^{n\delta_\kn}},
\]
which implies simultaneously for all $\w\in \Ttilde$
\[
\prod_{a \in \A} \tilde{q}(a|\w)^{\nwa}
\ge
\frac{
\prod_{a \in \A} \hat{q}(a|\w)^{\nwa}}
{2^{2|\A|^{\kn+1}\log n} \, 4^{n\delta_\kn}}.
\]
The above equation implies that $\tilde{q}$ and $\hat{q}$ are close distributions, since we can rearrange (and
divide both sides by $N_{\w}$) to obtain:
\begin{align*}
\sum_{a\in\cA} 
\frac{\nwa}{N_{\w}}
\log \frac{\hat{q}(a|\w)}{\tilde{q}(a|\w)}
&=
D\big(\hat{q}(\cdot|\w) \Vert \tilde{q}(\cdot|\w) \big)\\
&\le \frac{2 (|\A|^{\kn+1}\log n+n\delta_\kn)}{N_{\w}},
\end{align*}
where the first equality follows by writing out the value of the naive estimate, $\hat{q}(a|\w)={\nwa}/{N_{\w}}$.
\ignore{Since (see for example \cite{CT})
\[
D\big(\hat{q}(\cdot|\w) \Vert \tilde{q}(\cdot|\w) \big)
\ge 
\frac1{2\ln 2} 
\Vert\tilde{q}(\cdot|\w) -\hat{q}(\cdot|\w)\Vert_{_1}^2, 
\] 
we now have, with confidence bigger than $1-{2^{-|\A|^{\kn+1}\log n}}$, that for all $\w\in \Ttilde$ 
\[
\Vert\tilde{q}(\cdot|\w) -\hat{q}(\cdot|\w)\Vert_{_1} \le 2\sqrt{\frac{(\ln 2) (|\A|^{\kn+1}\log n+n\delta_\kn)}{N_{\w}}}.
\]
The Theorem follows from our Definition \ref{Good_States} of good states.}
\eTheorem
\bCorollary
Let $Y_1^n$ be generated by an unknown model $\ptq \in\cM_d$. Let $k_n=\alpha_n \log n $.
Given any $Y^0_{-\infty}$, with probability under $\ptq$ 
$\ge 1-\frac1{2^{|\A|^{\kn+1}\log n}}$, 
for all $\w\in \Ttilde$ simultaneously
\[
\Vert\tilde{q}(\cdot|\w) -\hat{q}(\cdot|\w)\Vert_{_1} \le 2\sqrt{\frac{(\ln 2) (|\A|^{\kn+1}\log n+n\delta_\kn)}{N_{\w}}}.
\]
\Proof
The proof follows immediately from Theorem \ref{thm:aggemptht} and Pinsker's inequality (see \eg \cite{CT}) 
\[
D\big(\hat{q}(\cdot|\w) \Vert \tilde{q}(\cdot|\w) \big)
\ge 
\frac1{2\ln 2} 
\Vert\tilde{q}(\cdot|\w) -\hat{q}(\cdot|\w)\Vert_{_1}^2. \eqed
\]
\label{CorL1dist}
\eCorollaryp

\ignore{\bCorollary
\label{CorGood}
Let $Y_1^n$ be generated by an unknown model $\ptq \in\cM_d$. Let $k_n=\alpha_n \log n $
and $\Gtilde$ be as Definition~\ref{Good_States}.
Given any $Y^0_{-\infty}$, with probability under $\ptq$ 
$\ge 1-\frac1{2^{|\A|^{\kn+1}\log n}}$, 
we have for all $\w\in \tilde{G}$ simultaneously
\[
\Vert\tilde{q}(.|\w) -\hat{q}(.|\w)\Vert_{_1} \le 2\sqrt{\frac{\ln 2}{\log n}+ \frac{\ln 2}{\log \frac{1}{\delta_{\kn}}  }  }.
\]
\Proof
The proof follows immediately from Corollary \ref{CorL1dist} and from our Definition \ref{Good_States} of good states.
\eCorollary}

\bRemark
We emphasize that the above results depend on $Y^0_{-\infty}$ hold for all history $Y^0_{-\infty}$. 
\eRemark

\bRemark
For all $\w \in \Ttilde$ with $|\A|=2$, by Corollary \ref{CorL1dist}, we have
\begin{align*}
\Vert\tilde{q}(.|\w) -\hat{q}(.|&\w)\Vert_{_1} \le \\
&\begin{cases} 2\sqrt{\frac{(2\ln 2) 2^{\kn+1}\log n}{N_{\w}}} 
& \textrm{ if $\kn \geq \log(\frac{n\delta_{\kn}}{2 \log n})$,} \\
      2\sqrt{\frac{(2\ln 2) n\delta_\kn}{N_{\w}}} 
      & \textrm{otherwise.} \\
   \end{cases}.
\end{align*}
If $d(i)=\gamma^i$ for some $0<\gamma<\frac{1}{2}$, then $\delta_\kn=\frac{\gamma^\kn}{1-\gamma}$ and by choosing $\kn=\gamma \log n$, the accuracy of above estimation is 
\[
\Theta\bigg(\sqrt{\frac{ n^{1+\gamma \log \gamma}}{N_{\w}}}\bigg).
\]
\ignore{\[
4\sqrt{\frac{(\ln 2) \sqrt{n}\log n}{N_{\w}}}.
\]}
Therefore if $N_\w=\frac{n}{\log n}$, then the accuracy of the estimation will be $\Theta\big(\sqrt{n^{\gamma\log \gamma}\log n}\big)$.
 
If $\frac{1}{2}<\gamma<1$, then the accuracy will be 
\[\Theta\bigg(\sqrt{\frac{n^\gamma\log n}{N_{\w}}}\bigg).\]
\ignore{\[
2\sqrt{\frac{(2\ln 2) n^{1+\frac{\log \gamma}{2}}}{(1-\gamma)N_{\w}}}.
\]} 
Therefore if $N_\w=\frac{n}{\log n}$, then the accuracy of the estimation will be $\Theta(n^{\frac{\gamma -1}{2}}\log n)$.

If $d(i)=\frac{1}{i^r}$ for some $r>2$, then $\delta_\kn\approx \kn^{1-r}$ and by choosing $\kn=\frac{1}{r}\log n-\log\log n$, the accuracy will be 
\[
\Theta\bigg(\sqrt{\frac{n}{\kn^{r-1} N_{\w}}}\bigg).
\]
\ignore{\[
2\sqrt{(2\ln 2) n/\kn N_{\w}}.
\]}
Therefore, if $N_\w=\frac{n}{\log n}$, then the accuracy of the estimation will be $\Theta\big(\sqrt{ \frac{1} {(\log n)^{r-2}}}\big)$.  

Good states $\Gtilde$ in Definition \eqref{Good_States} identify all strings $\w$ in the sample whose transition probabilities are accurate to at least $\frac{1}{\sqrt{\log n}}$. It is quite possible that for all $\w \in \Gtilde$, the accuracy is significantly better.
\eRemark

When the dependencies among strings die down exponentially, we can strengthen Theorem \ref{thm:aggemptht} with a more careful calculations to get a stronger convergence rate polynomial in $n$.
\bTheorem
\label{Exponential_Die}
Suppose $d(i)=\gamma^{i}$ for some $0<\gamma<1$.  Let $\zeta$ be a
nonnegative constant.
Let $k_n=\frac{\log n}{\log\frac{|\A|}{\gamma}}$. In analogy with $\tilde{G}$, we define
\[
\tilde{F}
\triangleq 
\bigg\{\w \in \Ttilde:\,
N_{\w}\geq n^{{}^{\zeta+\frac{\log|\A|}{\log |\A|-\log \gamma}}}
\bigg\}.
\]
Then, conditioned on any past $Y^0_{-\infty}$, 
with probability under $\ptq$ greater than
$1-{2^{-|\A|^{{}^{\kn+1}}\log n}} $ simultaneously
for all $\w \in \tilde{F}$ 
\begin{equation*}
\Vert\tilde{q}(.|\w) -\hat{q}(.|\w)\Vert_{_1} 
\le 
2 \sqrt \frac{\ln 2 \cdot \big((1-\gamma)|\A|\log n+1\big) }{(1-\gamma) n^{\zeta}}.
\end{equation*}
\Proof The proof of Theorem \ref{Exponential_Die} is similar to Theorem~{\ref{thm:aggemptht}}, but involves more careful but elementary algebra specific to the
exponential decay case using the value of $\kn$ and $\tilde{F}$ noted in the statement. Note that 
$\zeta 
< 
\frac
{\log\frac1 \gamma}
{\log\frac{|\A|}{\gamma}}$ for the Theorem not to be vacuously true.
\eTheorem
\bRemark
According to definition of good states in Theorem \ref{Exponential_Die} and the fact that $\Ttilde=\A^{k_n}$, we obtain 
\begin{equation*}
|\tilde{F}|\leq n^{-\frac{\log \gamma}{\log |\A |-\log \gamma}}, |\Ttilde|=n^{\frac{\log |\A|}{\log |\A |-\log \gamma}}
\end{equation*}
implying that if $\gamma \leq 1/|\A|$, all states of $\Ttilde$ can potentially be good.
\eRemark

\bRemark
 The rate of convergence in Theorem~\ref{thm:aggemptht} is the minimum that hold for all strings of
   length $\kn$ simultaneously, not just good strings-- and strings that appear more often will have stronger
  bounds automatically. Specifically, strings that appear $n^\beta$ times for any $\beta>0$ have
  convergence rates that are polynomial in $n$ for the exponential decay
  case. We emphasize that our results are not too off because even in \iid case, the Chernoff bounds do not provide much
  stronger relative to the bounds we obtain in the exponential decay case.
\eRemark 

\section{Estimation Along a Sequence of Stopping Times}
\label{stopping_times}

As we saw in the prior section, the aggregated parameters
$\tilde{q}(\cdot| \w)$ associated with any good state $\w \in
\tilde{G}$ can be approximated from the sample. From
Example~\ref{ergodic_ex}, we know that the stationary probabilities
may be a very sensitive function of the parameters associated with
states. \ignore{It is therefore perfectly possible that we estimate the
aggregated parameters at all strings in $\Gtilde$ reasonably well, but
are unable to gauge what the stationary probabilities of any string in
$\Gtilde$ may be.} How do we tell, therefore, if the few conditional
probabilities we estimated can say anything at all about stationary
probabilities? 

We explore these questions in the next two sections.  In order to
interpret the counts of various strings $\w$ in the sample, we first
study the process $\{Y_i\}_{i\geq 1}$ from $\ptq$ restricted to states in
$\tilde{G}$ in this section.  Some of the observations regarding
stopping times in this section are well known, see for
example~\cite{Nor98}.  

In Section~\ref{Dev_bound_Naive}, we use these observations with a
coupling argument to derive Theorem \ref{stat-estim} that provides
deviation bounds on the counts of strings. Note that any deviation
result must also keep in mind the fundamental difficulty that the
probability under the stationary distribution of all strings in
the observed sample could be arbitrarily small, no matter the size
of the sample, as illustrated in Example~\ref{arbit_stat_dist}.

\subsection{Restriction of $\ptq$ to $\Gtilde$}
To find deviation bounds for stationary distribution of good states,
we construct a new process $\{Z_m\}_{m\geq1}$, $Z_m\in \cT$
from the process $\{Y_i\}_{i\geq1}$. At the outset, note that $Z_m\in \cT$, where
$\cT$ is unknown. We use this process $\sets{Z_m}_{m\ge1}$ as an
analytical tool, and we will not need to actually observe it. We will
need to know if the process is aperiodic, but that can be resolved
by only looking at $\Gtilde$.

If $Y_{{i_{}}_{m}}$ is the
$(m+1)^{th}$ symbol in the sequence $\{Y_i\}_{i\geq1}$ such that
$\ctxs{\Ttilde}(Y_{-\infty}^{{i_{}}_{m}}) \in \tilde{G}$, then
$Z_{m}=\ctxs{\cT}(Y_{-\infty}^{{i_{}}_{m}})$.  The strong Markov
property \cite{Mey09} allows us to characterize $\sets{Z_m}_{m\geq1}$ as a
Markov process with transitions that are lower bounded by those
transitions of the process $\sets{Y_i}_{i\geq1}$ that can be
well estimated by the Theorems above.
More specifically, let 
\[
T_0 = \min\sets{ j\ge 0: \ctxtt(Y_{-\infty}^j)\in \tilde{G} },
\]
and let $Z_0= \ctxt(Y_{-\infty}^{T_0})$. For all $m\ge1$, $T_m$ is the
$(m+1)'$th occurrence of a good state in the sequence $\{Y_i\}_{i\geq1}$, namely
\[
T_m= \min\sets{ j>T_{m-1}: \ctxtt(Y_{-\infty}^{j})\in \tilde{G} },
\]
and $Z_m=\ctxt(Y^{T_m}_{-\infty})$. Note that $T_m$ is a \emph{stopping time}~\cite{LPW09}, 
and therefore $\sets{Z_m}_{m\geq1}$ is a Markov
chain by itself \footnote{Note that strong Markov property implies that the Markov property holds at stopping times.}. Let $\tilde{B}=\{ \s \in \Ttilde : \s \notin \tilde{G} \}$. 
The transitions between states $\w,\w'\in \tilde{G}$ are then the minimal, non-negative 
solution of the following set of equations in $\sets{\trp(\w|\s): \s\in \A^{k_n}, \w\in \tilde{G}}$
\begin{align*}
\trp(\w | \w') &= \ptq\Paren{\ctxtt(Y_{-\infty}^1)=\w|\ctxtt(Y_{-\infty}^{0})=\w'}\\
&+ 
\sum_{\s\in \tilde{B}}
\trp(\w | \s) 
\ptq\Paren{\ctxtt(Y_{-\infty}^{1})=\s|\ctxtt(Y_{-\infty}^{0})=\w'}.
\end{align*}
An important point to note here is that if $\w$ and $\w'$ are good states, 
\[
\trp(\w | \w')
\ge
\ptq\Paren{\ctxtt(Y_{-\infty}^{1})=\w|\ctxtt(Y_{-\infty}^{0})=\w'},
\]
and the lower bound above can be well estimated from the sample 
as shown in Theorem~\ref{thm:aggemptht}.
\bDefinition
We will call $\{Z_m\}_{m \geq 1}$, the restriction of $\ptq$ to $\tilde{G}$. 
\eDefinition

\subsection{Properties of $\sets{Z_m}_{m\ge1}$}
\bProperty A few properties about $\{Z_m\}_{m \geq 1}$ are in order. $\{Z_m\}_{m \geq 1}$ 
is constructed from an irreducible process $\{Y_i\}_{i\geq1}$, thus $\{Z_m\}_{m \geq 1}$ is irreducible as well. Since $\{Y_i\}_{i\geq1}$ is 
positive recurrent, so is $\{Z_m\}_{m \geq 1}$. But despite $\{Y_i\}_{i\geq1}$ being aperiodic, $\sets{Z_m}_{m \geq 1}$ could be periodic as in the Example below. But periodicity of 
$\sets{Z_m}_{m \geq 1}$ can be determined by $\Gtilde$ alone (because $\cT$, while unknown, is a full, finite $\cA$-ary tree). 
\label{prop:two}
\eProperty

\bExample Let $\{Y_i\}_{i\geq1}$ be a process generated by
context tree model $\ptq$ with $\T=\{11,01,10,00\}$ and
$q(1|11)=\frac{1}{2},q(1|01)=\epsilon,q(1|10)=1-\epsilon,q(1|00)=\frac{1}{2}$. If
$\epsilon>0$, then $\ptq$ represents a stationary aperiodic Markov
process. If $\{Z_m\}_{m\geq1}$ be the restriction of process $\{Y_i\}_{i\geq1}$ to
$\Gtilde=\{01,10\}$, the restricted process will be 
periodic with period 2.  \eExample

\bProperty
\label{prop2}
Suppose $\sets{Z_m}_{m\geq 1}$ is aperiodic. 
Let $\stn_Y$ and $\stn_Z$ denote the stationary distribution of the processes $\{Y_i\}_{i\geq1}$ 
and $\{Z_m\}_{m\geq 1}$, respectively, with $n$ samples of a sequence $\{Y_i\}_{i\geq1}$ yielding $m_n$
samples of $\sets{Z_m}_{m\geq 1}$. Similarly, let $\stn_Z(\w)$ denote the stationary probability
of the event that $\w$ is a suffix of samples in the process $\{Z_m\}_{m\geq 1}$. Then for all $\w,\w' \in \tilde{G}$ (note that $\stn_Y(\w')> 0$)
\begin{align*}
\frac{\stn_Y(\w)}{\stn_Y(\w')}
&\aeq{wp 1}
\frac{ \lim\limits_{n \to \infty} \sum_{i=1}^{n} \frac{\indctr(\ctxtt(Y_{-\infty}^i)=\w)} {n} }  
{\lim\limits_{n \to \infty} \sum_{i=1}^{n} \frac{\indctr(\ctxtt(Y_{-\infty}^i)=\w')} {n} }\\
&=
\frac{ \lim\limits_{m_n \to \infty} \sum_{j=1}^{m_n} \frac{\indctr(\w\preceq Z_j)} {m_n} }  
{\lim\limits_{m_n \to \infty} \sum_{j=1}^{m_n} \frac{\indctr(\w'\preceq Z_j)} {m_n} } \\
&\aeq{wp 1}
\frac{\stn_Z(\w)}{\stn_Z(\w')}. 
\end{align*}
\eProperty

\section{Estimate Of Stationary Probabilities}
\label{Dev_bound_Naive}
Thus far, we have identified a set $\Gtilde\subseteq \cA^{\kn}$ of
good strings using $n$ observations from $\ptq$.
For strings $\w\in\Gtilde$, we have been able to estimate approximately
the conditional distributions,
conditioned on past strings $\w$---namely $\prob(Y|\w)$, or
equivalently the aggregated parameters
$\tilde{q}(\cdot|\w)$. As mentioned before, it is not clear if this information we have obtained from the sample will
allow us to say anything at all about the stationary probabilities 
of strings $\w\in\Gtilde$. This section develops on this question, and 
shows how to interpret naive counts of $\w\in\Gtilde$ in the sample.

\subsection{Outline}
Our main result in this section is Theorem~\ref{stat-estim}, a concentration
result on the counts $N_{\w}$. This follows essentially as a martingale
convergence bound. Consider the process restricted to the good states $\Gtilde$, namely, 
$\{Z_i\}_{i\geq 0}$. In Definition~\ref{dfn:stn}, we consider the natural Doob
martingale $V_m \ed \E[ N_{\w} |Z_0,Z_1\upto Z_m]$.
In Subsection~\ref{s:stn:cpl}, we give a bound on the differences $|V_m- V_{m+1}|$
using a variation of the coupling argument. In Subsection~\ref{coup_desc}, we construct
a specific coupling that is analyzed in~\ref{Mart_diff} to provide a bound
on the martingale differences. Theorem~\ref{stat-estim} then follows from Azuma's
inequality.
\subsection{Preliminaries}

\newcommand{\osym}{a}
For any (good) state $\w$, let $G_\w\subset \cA$ be the set of letters
that take $\w$ to another good state,
\begin{equation} 
\label{eq:gu}
G_\w = \sets{\osym \in\cA: \ctxs{\Ttilde}(\w\osym )\in \Gtilde}.
\end{equation}
Our confidence in the empirical
counts of good states matching their (aggregated) stationary probabilities 
follows from a coupling argument, and depends on the following parameter
\begin{equation}
\label{eq:cplngg}
\cplngg 
= 
\min_{\U,\V\in \tilde{G}} 
\sum_{\substack{\osym\in G_\U\cap G_\V}} \min \sets{{\tilde q}(\osym|\U), {\tilde q}(\osym|\V) },
\end{equation}
where $\tilde{q}(\osym|\U)$ is as defined before in~(\ref{trans_agg}) for aggregated model parameters.
The parameter $\cplngg$ is closely related to the notion of Dobrushin's ergodicity coefficient of Markov processes (see e.g., \cite{GMS08}). The parameter $\cplngg$ is different from standard Dobrushin's ergodicity coefficient in two aspects. First, notice that in \eqref{eq:cplngg}, $\U$ and $\V$ might not be the states of $\ptq$. Second, the summation index runs over those alphabets which take both states $\U$ and $\V$ to any other good state in $\tilde{G}$.  
Note that for any state $\s\in\T$ of the original process $\ptq$, if $\U\preceq\s$, 
using Proposition \ref{die_out}
\begin{equation}
\tilde{q}(\osym|\U) 
\leq \frac{q(\osym|\s)}{1-\delta_{|\U|}}.
\label{eq:q_u}
\end{equation}
\bRemark Recall that the deviation bounds in
Theorems~\ref{thm:aggemptht} and~\ref{Exponential_Die} hold
simultaneously for all contexts in $\Gtilde$. The above definition
only depends on parameters associated with $\Gtilde$. Hence we can
estimate $\cplngg$ from the sample with the same confidence and
(twice the) error given in those Theorems.
\eRemark

The counts of various $\w\in\Gtilde$ then concentrates as shown in the
Theorem~\ref{stat-estim}, and how good the concentration is can be estimated as a
function of $\cplngg$ (and $\delta_\kn$) and the total number of times 
states in $\Gtilde$ occur in the sample. Now $\Gtilde$ as well as $\cplngg$ are
well estimated from the sample---thus we can look at the data to
interpret the empirical counts of various substrings of the data.

Let
\[
\Delta_j=\sum_{i\geq j} \delta_i.
\]
For the following results, we require
$\{\delta_i\}_{i \geq 1}$ to be summable. Thus, $\Delta_j$ is finite for all $j$ and decreases
to 0 as $j$ increases. If $d(i)\sim \gamma^i$, then $\Delta_j$ also
diminishes as $\gamma^j$. But if $d(i)\sim \frac1{i^r}$ diminishes
polynomially, then $\Delta_j$ diminishes as $1/j^{r-2}$. If
$d(i)=1/i^{2+\eta}$ for any $\eta>0$, we therefore satisfy the
summability of $\{\delta_i\}_{i \geq 1}$.  However, $d(i)$ can also diminish as
$1/(i^2\poly(\log i))$ for appropriate polynomials of $\log i$
for the counts of good states to converge. In what follows, we assume that $\delta_{i} \leq \frac{1}{i}$.

\bDefinition 
\label{dfn:stn}
Let $\Gtilde$ be the set of good states from Definition
\ref{Good_States}.  Let $\ntilde$ be total count of all good states in
the sample and $\ell_n$ denote the smallest integer such that
$\Delta_{\ell_n} \leq \frac{1}{n}$. To analyze the naive counts of $\w
\in \tilde{G}$, we define
\begin{equation*}
V_m \ed \E[ N_{\w} |Z_0,Z_1\upto Z_m],
\end{equation*}
where $N_{\w}$ is as in Definition \ref{impr_theta}.  
\eDefinition

Observe that $\sets{ V_m}_{m=0}^\ntilde$ is a Doob martingale. Note that
\[
V_0= \E [N_\w|Z_0] \text{ and }V_\ntilde= N_\w.
\]  
Once again, note that we do not have to observe the restriction process 
$\sets{Z_m}_{m\ge1}$ at any point, nor do we have to observe the martingale
$V$ except for noting $V_\ntilde=N_\w$.
\bRemark To prove Theorem~\ref{stat-estim}, we first bound the differences $|V_{m}-V_{m-1}|$ of the martingale
using a coupling argument in Lemma~\ref{Doob_bound}. 
Since the memory of the process $\ptq$ could be large, our coupled chains may not actually coalesce in the usual sense. But they get ``close enough''
that the chance they diverge again within $n$
samples is less than $1/n$. Once we bound the differences in the martingale
$\sets{V_m}_{m=0}^\ntilde$, Theorem~\ref{stat-estim} follows as an easy application of
Azuma's inequality. 
\eRemark

\subsection{The Coupling Argument}
\label{s:stn:cpl}
Since for all $m\ge 1$
\begin{align*}
|V_m- & V_{m-1}|
= 
\left|
\E[ N_{\w} | Z_0\upto Z_m] 
- 
\E[ N_{\w} | Z_0\upto Z_{m-1}]
\right|\\
\ignore{&=
\left|
\E[ N_{\w} | Z_0\upto Z_m] 
- 
\E_{Z_im}
\E[ N_{\w} | Z_0\upto Z_{m-1}, Z_m]
\right|\\}
&\le
\max_{Z_m^{'},Z_m^{''}}
\left|
\E[ N_{\w} | Z_0\upto Z_m^{'}] 
- 
\E[ N_{\w} | Z_0\upto Z_m^{''}]
\right|,
\end{align*}
we bound the maximum change in $N_{\w}$ if the $m$th good state was
changed into another (good) state. 

Suppose there are sequences $\{Z^{'}_j\}_{j= m}^n$ (starting from
state $Z_m^{'}$) and $\{Z^{''}_j\}_{j=m}^n$ (starting from state
$Z_m^{''}$), both faithful copies of the restriction of $\ptq$ to
$\Gtilde$ but coupled with a joint distribution $\coplg$ to be
described below. From the coupling argument of Section~\ref{s:b:cpl},
we have for $\w\in\Gtilde$ (hence $|\w|=\kn$) for all $\coplg$
\begin{align}
\label{eq:aldous}
|\E[ N_{\w} | Z_0\upto Z_m^{'}] - \E[ N_{\w} | & Z_0\upto Z_m^{''}]| \nonumber\\
&\le 
\sum_{j=m+1}^n \coplg(\Zo_j\napproxkn \Zt_j),
\end{align}
where we use
\[
\Zo_j\approxkn \Zt_j  \text{ for }\ctxs{\cA^{\kn}}(\Zo_j)=\ctxs{\cA^{\kn}}(\Zt_j).
\]
We will bound the right side of~\eqref{eq:aldous} above using 
properties of $\coplg$ that we describe. Note that the summation goes up to $n$ since no matter what 
$Z_m^{'}$ and $Z_m^{''}$ are, a length-$n$ sample can have at most $n$ good states on the right side of \eqref{eq:aldous}.
The coupling technique is a convenient ``thought experiment'' that
culminates in Theorem~\ref{stat-estim} giving a deviation bound on the
naive counts of states. We do not actually have to generate the two
chains as part of the estimation algorithm, nor do we need to observe
the martingale $V$, except for noting $V_\ntilde=N_\w$.

In Subsection~\ref{coup_desc} below, we describe $\coplg$,
highlight some properties of the described coupling in Subsection~\ref{coupling_prop},
and use the said properties to obtain a bound on the martingale difference
in Subsection~\ref{Mart_diff}.

\subsection{Description of Coupling $\coplg$}
\label{coup_desc}
Suppose we have $Z^{'}_{j}$ and $Z^{''}_{j}$.  This subsection
describes how to obtain next sample $Z^{'}_{j+1}$ and $Z^{''}_{j+1}$
of the two coupled chains, namely how to sample from
\[
\coplg(Z^{'}_{j+1},Z^{''}_{j+1}| Z^{'}_{j},Z^{''}_{j}).
\]
Recall from section \ref{s:b:cpl} that individually taken, both
$Z^{'}_{j+1}$ and $Z^{''}_{j+1}$ are faithful evolutions of the
restriction of $\ptq$ to $\tilde{G}$.
However, given $Z^{'}_{j}$ and $Z^{''}_{j}$, $Z^{'}_{j+1}$ and $Z^{''}_{j+1}$
are not necessarily independent.

To obtain $\Zo_{j+1}$ and $\Zt_{j+1}$, starting from states $\Zo_{j}$ and
$\Zt_j$ we run copies $\sYo$ and $\sYt$\footnote{Note that $\sYo$ and $\sYt$ are sequences of symbols from $\cA$, generated according to transitions defined by $\ptq$.}of coupled chains individually
faithful to $\ptq$. If $l$ is the smallest number such that 
\[
\ctxtt(Z^{'}_{j} Y^{'}_{j1}\cdots Y^{'}_{jl}) \in \tilde{G},
\] 
then 
\[
\Zo_{j+1}=\ctxt(Z^{'}_{j} Y^{'}_{j1}\cdots Y^{'}_{jl}).
\] 
Similarly for $Z^{''}_{j+1}$.
\subsubsection{\underline{Sampling from the joint distribution $\coplg(Y^{'}_{j1},Y^{''}_{j1}| \Zo_j,\Zt_j)$}}\tcw{dummy}\\

\begin{figure}[!t]
\centering
\includegraphics[trim=4cm 11cm 3cm 10.5cm, clip,scale=.65]{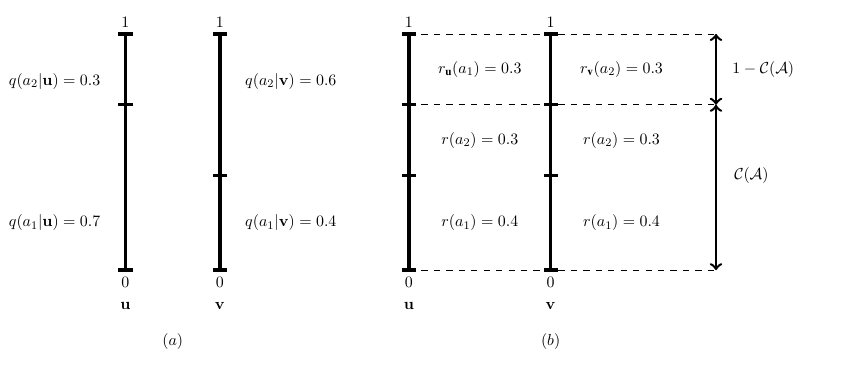}
\caption{(a) The conditional probabilities with which $\Yos1$ and
  $\Yts1$ have to be chosen respectively are $q(\cdot|\U)$ and
  $q(\cdot|\V)$. The line on the left determines the choice of $\Yos1$
  and the one on the right the choice of $\Yts1$. For example, if
  $U_{j1}$ is chosen uniformly in [0,1], the probability of choosing
  $\Yos1=a_1$ is $q(a_1|\U)$. Instead of choosing $\Yos1$ and $\Yts1$
  independently, we will reorganize the intervals in the lines so as
  to encourage $\Yos1=\Yts1$. (b) Reorganizing the interval $[0,1]$
  according to the described construction. Here
  $r(a_1)=\min\sets{q(a_1|\U), q(a_1|\V)}$ and similarly for
  $r(a_2)$. If $U_{j1}$ falls in the interval corresponding to
  $r(a_1)$, then $(\Yos1,\Yts1)= (a_1,a_1)$. If $U_{j1}>C(\cA)$ in
  this example, then $(\Yos1,\Yts1)=(a_1,a_2)$.  When $U_{j1}$ is
  chosen uniformly in [0,1], the probability $\Yos1$ outputs any
  symbol is the same as in the picture on the left, similarly for
  $\Yos2$. }
\label{interval}
\end{figure}

While the following description appears verbose, Fig. \ref{interval}. (b) represents the description pictorially. Specifically, the chains $\sYo$ and $\sYt$ are coupled as follows. 
We generate a number $U_{j1}$ uniformly distributed in $[0,1]$. Given
$(\Zo_j \text{ and } \Zt_j)$ with suffixes $\U$ and $\V$ respectively
in $\Gtilde$, we let $G_\U\in\cA$ (and $G_\V$ similarly) be the set of
symbols in $\cA$ defined as in~\eqref{eq:gu}. We split the interval
from 0 to 1 as follows: for all $a\in\cA$, we assign intervals $r(a)$
of length $\min \sets{q(a|\U), q(a|\V) }$, in the following order:
we first stack the above intervals corresponding to $a\in G_\U\cap
G_\V$ (in any order) starting from 0, and then we put in the intervals
corresponding to all other symbols. Now let,
\[
(\Yos1, \Yts1) 
= 
(a,a) \text{ if } U_{j1} \in r(a).
\]
Let 
\begin{equation}
\label{eq:ca}
C(\cA)=\sum_{a'\in\cA} r(a')=\sum_{a'\in\cA} \min\sets{q(a'|\Zo_j), q(a'|\Zt_j) },
\end{equation}
be the part of the interval is already filled up. Thus if $U_{j1}<C(\cA)$, equivalently with probability $C(\cA)$, the
two chains output the same symbol. We use the rest of the interval $[C(A),1]$ in any valid way to satisfy the fact that
$\Yos1$ is distributed as $\ptq(\cdot|\Zo_j)$ and $\Yts1$ is distributed as $\ptq(\cdot|\Zt_j)$. For one standard
approach, for all $a$ assign
\[
r_\U(a)= (q(a|\U)- q(a|\V))^+ = \max\sets{ q(a|\U)-q(a|\V), 0}
\]
and similarly $r_\V(a)$. 
Note that only one of $r_\U(a)$ and $r_\V(a)$ can be strictly positive and that for all $a$, $r(a)+r_\U(a)=q(a|\U)$ while
$r(a)+r_\V(a)=q(a|\V)$. Therefore,
\[
\sum_{a'\in\cA} r_\U(a') = \sum_{a'\in\cA}r_\V(a') =1-C(\cA).
\] 
An example of such construction for binary alphabet is illustrated in Fig. \ref{interval} in which
we have assumed $G_\U\cap G_\V=\{a_1\}$.
We will keep two copies of the interval $[C(\cA),1]$, and if $U_{j1}>C(\cA)$ we output $(\Yos1, \Yts1)$ based on where
$U_{j1}$ falls in both copies. We will stack the first copy of $[C(\cA),1]$ with intervals of length $r_\U(a)$ for all $a$
and the second copy of $[C(\cA),1]$ with intervals length $r_\V(a)$ for all $a$.  We say $U_{j1} \in (r_\U(a),r_\V(a'))$ if
$U_{j1}\in r_\U(a)$ in the first copy and $U_{j1}\in r_\V(a')$ in the second copy,
\[
(\Yos1, \Yts1) 
= 
(a,a') \text{ if } U_{j1}\in (r_\U(a),r_\V(a')).
\]
Note in particular that
\[
\coplg(\Yos1| \Zo_j, \Zt_j)= \ptq(\Yos1|\Zo_j).
\]
and similarly for $\Yts1$.
If $\ctxs{\Ttilde}(\Zo_j\Yos1)\in\Gtilde$ and
$\ctxs{\Ttilde}(\Zt_j\Yts1)\in\Gtilde$, we have $\Zo_{j+1}$ and
$\Zt_{j+1}$. There is therefore no need for further samples $\Yos2$
and $\Yts2$ onwards. 
\subsubsection{\underline{Sampling from $\coplg(\{Y^{'}_{ji},Y^{''}_{ji}\}_{i \geq 2}| \Zo_j \Yos1, \Zt_j\Yts1)$}}\tcw{dummy}\\

In case at least one of the following holds:
$\ctxs{\Ttilde}(\Zo_j\Yos1)\notin\Gtilde$ or
$\ctxs{\Ttilde}(\Zt_j\Yts1)\notin\Gtilde$, then the following
subsection explains how to proceed. 
\begin{enumerate}
\item If $\Yos1=\Yts1$ but only one of $\ctxs{\Ttilde}(\Zo_j\Yos1)\in\Gtilde$ and
  $\ctxs{\Ttilde}(\Zt_j\Yts1)\in\Gtilde$, then we have one of $\Zo_{j+1}$ and $\Zt_{j+1}$. To get the other, we
  continue (according to transitions defined by $\ptq$) only its corresponding chain till we get a good state.
\item If $\Yos1=\Yts1$, $\ctxs{\Ttilde}(\Zo_j\Yos1)\notin\Gtilde$ and $\ctxs{\Ttilde}(\Zt_j\Yts1)\notin\Gtilde$, we need
  to continue both chains. We generate $\Yos2, \Yts2$ as we did for the first samples---by generating a new random
  number $U_{j2}$ uniform in $[0,1]$, and by coupling as in Fig. \ref{interval}. (b) and the two distributions $q(\cdot|\Zo_j \Yos1)$ and
  $q(\cdot|\Zt_j \Yts1)$. And continue in this fashion so long as the samples in the two chains remain
  equal but do not hit good contexts. This will be case that will be most important for us later on.
\item If $\Yos{l}\ne\Yts{l}$ at any point and neither chain has seen a good state yet, we just run the chains
  independently from that point on for how long it takes each to hit a good
  aggregated state.
\end{enumerate}
Once again we have
\begin{align*}
\coplg(\Yos{(r+1)}| \Zo_j \Yos1\upto\Yos{r}, & \Zt_j\Yts1\upto\Yts{r})
\\&= \ptq(\Yos{(r+1)}|\Zo_j \Yos1\upto\Yos{r}),
\end{align*}
and similarly for $\Yts{(r+1)}$.

\subsection{Some Observations on Coupling}
\label{coupling_prop}
For any $r$, let $\Zo_{r} \sim \Zt_{r}$ denote the following event that happens to be a subset of case where we do not need $\Yoss{r}2$ and $\Ytss{r}2$ onwards,
\[
\Sets{ 
\Yoss{r}1= \Ytss{r}1
\text{ and } 
\ctxs{\Ttilde}(\Zo_{r}\Yoss{r}1)\in\Gtilde \text{ and } \ctxs{\Ttilde}(\Zt_{r}\Ytss{r}1)\in\Gtilde}.
\]
Recall the definition of $\cplngg$ from~\eqref{eq:cplngg}.

\noindent
\bObservation
\label{eq:one}
$\coplg(\Zo_{r}\sim \Zt_{r}| \Zo_{r-1}, \Zt_{r-1}) \ge \cplngg(1-\delta_\kn)$.
\Proof
Combining~\eqref{eq:cplngg} and~\eqref{eq:q_u}, we have
\begin{align*}
\coplg(\Zo_{r}\sim \Zt_{r}| \Zo_{r-1}, \Zt_{r-1})
&=
\sum_{a\in\cA} \min\sets{q(a|\Zo_{r-1}), q(a|\Zt_{r-1}) } \\
&\ge
\cplngg(1-\delta_\kn).
\end{align*}
\eObservation

\ignore{where the $1-\delta_\kn$ term comes because the parameter $\cplngg$ is defined on the aggregated
parameters, but $\Yo$ and $\Yt$ evolve according to $\ptq$.}
\noindent
Furthermore, if $\Zo_{i}\sim \Zt_{i}$ for the $\kn$ consecutive samples, $j-\kn+1\le i\le j$, then we have
\[
\Zo_{j}\approxkn \Zt_{j}. 
\]
To proceed, once $\Zo_{j}\approxkn\Zt_{j}$, we would like the two chains to coalesce tighter in every subsequent step,
namely we want for all $1\le l\le n$, $\Zo_{j+l}\approxs{\kn+l} \Zt_{j+l}$.  Starting from $\Zo_{j}\approxkn\Zt_{j}$,
we can have $\Zo_{j+1}\approxs{\kn+1}\Zt_{j+1}$ if 
\begin{enumerate}
\item $\Zo_{j+1}\sim\Zt_{j+1}$, or 
\item if the chains $\sYo$ and $\sYt$ evolve through a sequence of $m>1$ steps before hitting a context in $\Gtilde$ on
  the $m'$th step with $\Yos{l}= \Yts{l}$ for each $l\le m$.
\end{enumerate}

\bObservation
\label{obs_coels}
Suppose $\Zo_j\approxkn \Zt_j$. While sampling from $\coplg$ for the samples $\Zo_{j+1}$ and $\Zt_{j+1}$, suppose 
$\Yos{i}=\Yts{i}$ for $i\ge1$. If $m$ is the first time the first chain 
hits a good context, namely $m$ is the smallest number such that
\[
\ctxs{\cA^\kn}(Z^{'}_{j}\{Y^{'}_{ji} \}_{i=1}^{m}) \in \Gtilde,
\]
it follows that the second chain also hits a good context at the same time, namely
\[
\ctxs{\cA^\kn}(Z^{''}_{j}\{Y^{'}_{ji} \}_{i=1}^{m})
= 
\ctxs{\cA^\kn}(Z^{''}_{j}\{Y^{''}_{ji} \}_{i=1}^{m})
\in \Gtilde.
\]
Note that we may not be able to say the above if $\Zo_j\napproxkn \Zt_j$. Furthermore, now we also have $\Zo_{j+1}\approxs{\kn+m} \Zt_{j+1}$. 
\eObservation

\ignore{
This is the situation in case 3 of the list above,
but in addition in each step $l$,
\[
\ctxs{\cA^\kn}(Z^{'}_{j}\{Y^{'}_{ji} \}_{i=1}^{l}) = 
\ctxs{\cA^\kn}(Z^{''}_{j}\{Y^{''}_{ji} \}_{i=1}^{l}),
\]
since $\Zo_{j}\approxkn\Zt_{j}$. Therefore, both chains will hit a common good context in $\Gtilde$ in 
$m$ steps. But in addition we will also have
\[
\Zo_{j+1}\approxs{\kn+m}\Zt_{j+1}.
\] }

\noindent
Let us now bound how likely this sort of increasingly tighter merging is.
Because of the way we have set up our coupling, the probability 
\begin{align*}
\coplg(\Yos1=\Yts1 |\Zo_{j}\approxkn\Zt_{j} ) 
&=
\sum_{a\in\cA} \min\Sets{ q(a|\Zo_{j}), q(a|\Zt_{j})}\\
&\ge
\sum_{a\in\cA} \tilde{q}\Paren{a|\ctxtt\paren{\Zo_{j}}}(1-\delta_\kn)\\
&= 1-\delta_\kn,
\end{align*}
where $q$ and $\tilde{q}$ are the conditional probabilities associated with $\ptq$ and $\ptildettildeq$ respectively. Similarly 
\begin{align*}
\coplg\big(
\Yos{(l+1)}=\Yts{(l+1)}
\big|\Zo_{j}\approxkn\Zt_{j}, \{\Yos{i}\}_{i=1}^{l}&=\{\Yts{i}\}_{i=1}^{l}\big)\\
&\ge 1-\delta_{\kn+l}.
\end{align*}
It is important to note that the above statement holds whether
\[
\ctxs{\cA^\kn}(\Zo_j \sets{\Yos{i}}_{i=1}^l)\in\Gtilde
\text{ or }
\ctxs{\cA^\kn}(\Zo_j \sets{\Yos{i}}_{i=1}^l)\notin\Gtilde,
\]
and is simply a consequence of dependencies dying down.
Therefore (no matter what $m$ is),
\[
\coplg(\Zo_{j}\approxs{\kn+1} \Zt_{j}| 
\Zo_{j-1}\approxkn\Zt_{j-1}) 
\ge \prod_{l=\kn}^\infty (1-\delta_l) \stackrel{(i)}{\ge} 1-\Delta_\kn,
\]
where the inequality in $(i)$ is by Proposition \ref{Bonferroni}.

The above equation, while accurate, is not the strongest we can say with essentially the same argument.
We are now going to progressively make stronger statements with the same arguments.
First, note that we obtain for all $\ell$, the event 
\[
\sets{\exists\, \ell' \le \ell \text{ s.t. }\Zo_{j+\ell'}\approxs{\kn+\ell}\Zt_{j+\ell'}},
\] 
can happen by going through a sequence of tighter and tighter coalesced
transitions of $\ptq$ (no matter in how many of those steps we saw contexts in
$\Gtilde$). Therefore,
\begin{align*}
\coplg(\exists\, \ell' \le \ell \text{ s.t. }\Zo_{j+\ell'}\approxs{\kn+\ell}\Zt_{j+\ell'}| \Zo_j\approxkn\Zt_j)
&\ge 
\prod_{l=\kn}^{\infty} (1-\delta_l)\\
&\ge 1-\Delta_\kn.
\end{align*}
And we can easily strengthen the above to say for all $\ell$,
\begin{equation*}
\coplg(\Zo_{j+\ell}\approxs{\kn+\ell}\Zt_{j+\ell}| \Zo_j\approxkn\Zt_j)
\ge 
1-\Delta_\kn,
\end{equation*}
for the same reason. Indeed, we can further strengthen the above statement to note: 
\bObservation
\label{eq:strng}
If $\Zo_j\approxs{\ell}\Zt_j$ for any $\ell\ge\kn$, the chance of ever diverging,
\[
\coplg(\exists l>0 \text{ s.t. }\Zo_{j+l}\napproxs{\ell}\Zt_{j+l}| \Zo_j\approxs{\ell}\Zt_j)
\le \Delta_{\ell}.
\eqed
\]
\eObservationp
This motivates following definition of coalescence when dealing with finite length-$n$ samples.
\bDefinition
\label{dfn:coa}
We say the chains $\sZo_{i \geq 1}$ and $\sZt_{i \geq 1}$ have \emph{coalesced} if for any $j$, 
\[\Zo_j\approxs{\max\sets{\kn,\ell_n}}\Zt_j,
\]
where as in Definition~\ref{dfn:stn}, $\ell_n$ is the smallest number satisfying $\Delta_{\ell_n}\le1/n$.
Note that our definition of coalescence differs from regular literature which requires
the equality in above definition. 
\eDefinition
\subsection{Deviation bounds on stationary probabilities}
\label{Mart_diff}

\bDefinition
\label{eq:cb}
Let
\[
\cB
\ed
\cB(\kn,\ell_n,\cplngg)
\ed
1+\frac{4\max\sets{\ell_n,\kn} }{\cplngg^\kn(1-\Delta_\kn)}, 
\]
where $\ell_n$ is as in Definition~\ref{dfn:coa} and $\cplngg$ is 
from~\eqref{eq:cplngg}.
\eDefinition

\bRemark
The above quantity $\cB$ will be a determining factor in how well we
can estimate stationary probabilities. At this point, it may be useful
to estimate the order of magnitude of the various terms. First,
$\kn=\alpha_n\log n$ as mentioned before. The parameter
$\cplngg<1$. The results below hold when $\cplngg$ is not too small,
it should be a constant such that
$\cplngg^\kn=\omega(1/\sqrt{n})$. Since $\kn=\alpha_n\log n$, we
typically can take $\alpha_n$ as large as a small constant in light of
the above.

While the results below do not require it, it is helpful to keep the
case $\tilde{n}=\cO(n)$ in mind while interpreting the results below.

Note that if the dependencies die
down exponentially, namely $d(i)=\gamma^i$ for some $ 0< \gamma < 1$,
then $\ell_n = \ceil{\log \big(n/(1-\gamma)^2 \big)/\log
  1/\gamma}$. If the dependencies die down polynomially, namely
$d(i)=1/i^r$ for some $r>2$, then $\ell_{n} =
\ceil{2+\Paren{\frac{n}{(r-1)(r-2)}}^{\frac1{r-2}}}$. 
\eRemark

\bLemma
\label{Doob_bound}
Let $\sets{V_m}_{m=0}^\ntilde$ be the Doob martingale described in Definition~\ref{dfn:stn}. For all $m\geq 1$,
\[
|V_m-V_{m-1}|
\leq
\cB(\kn,\ell_n,\cplngg),
\]
where $\cB(\kn,\ell_n,\cplngg)$ is as defined in Definition~\ref{eq:cb}.
\Proof
Recall that 
\begin{align*}
|V_m-& V_{m-1}|
= 
\left  |
\E[ N_{\w} | Z_0\upto Z_m] 
- 
\E[ N_{\w} | Z_0\upto Z_{m-1}]
\right  |\\
\ignore{&=
\left|
\E[ N_{\w} | Z_0\upto Z_m] 
- 
\E_{Z_im}
\E[ N_{\w} | Z_0\upto Z_{m-1}, Z_m]
\right|\\}
&\le
\max_{Z_m^{'},Z_m^{''}}
\left|
\E[ N_{\w} | Z_0\upto Z_m^{'}] 
- 
\E[ N_{\w} | Z_0\upto Z_m^{''}]
\right|\\
&\le
\sum_{j=m+1}^n \coplg(\Zo_j\napproxkn \Zt_j).
\end{align*}
Let $\tau$ be the smallest number bigger than $m$ such that $\Zo_\tau\approxs{\ell_n}\Zt_\tau$.
For any positive integer $t'$, the probability $\tau>t'\ell_n$ can be upper bounded
by splitting the $t'\ell_n$ samples of the two chains
$\sets{\Zo_j}_{j=m+1}^{m+t'\ell_n}$ and $\sets{\Zt_j}_{j=m+1}^{m+t'\ell_n}$ into blocks of length $\ell_n$. While the approach remains the same, we consider two calculations: (i) $\kn\leq \ell_n$ and (ii) $\kn \ge \ell_n$ below.

If $k_n \leq \ell_n$, the probability the two chains coalesce in any
single block is, using Observation~\ref{obs_coels} for $\kn$ times and
then Observation~\ref{eq:strng}
\begin{align*}
\ge & \,\cplngg^\kn(1-{\delta_{\kn}})^\kn (1-\Delta_\kn) \\
\stackrel{(i)}{\ge}& \, \cplngg^\kn (1-\frac{1}{\kn})^\kn (1-\Delta_\kn)\\
\ge &\, \cplngg^\kn(1-\Delta_\kn)/4,
\end{align*}
where $(i)$ is because $\delta_\kn \leq \frac{1}{\kn}$.

Thus, 
\[
\coplg(\tau>t'\ell_n)\le \Paren{1-\cplngg^\kn(1-\Delta_\kn)/4}^{t'}.
\]
Furthermore, $\E\tau-m$ can be bounded using the expected number of blocks
before the chains merge in any single block, thus,
\[
\E\tau \le m+ \frac{4\ell_n}{\cplngg^\kn(1-\Delta_\kn)}.
\]
Then for all $j\ge m+1$
\begin{align*}
\coplg(\Zo_j\napproxs{\ell_n} \Zt_j)
&= 
\coplg(\Zo_j\napproxs{\ell_n} \Zt_j \text{ and } \tau < j)
+
\coplg(\Zo_j\napproxs{\ell_n} \Zt_j \text{ and } \tau \ge j)\\
&\ale{(i)}
\Delta_{\ell_n}+\coplg(\Zo_j\napproxs{\ell_n} \Zt_j \text{ and } \tau \geq j)\\
&\le 
\frac1n+\coplg(\tau \ge j),
\end{align*}
where inequality $(i)$ above follows because $\Zo_\tau\approxs{\ell_n}\Zt_\tau$ by definition
and from Observation~\ref{eq:strng}.
Finally we upper bound~\eqref{eq:aldous},
\begin{align}
\nonumber
\sum_{j=m+1}^n \coplg(\Zo_j\napproxkn \Zt_j)
&\le 
\sum_{j=m+1}^n\coplg(\Zo_j\napproxs{\ell_n} \Zt_j)\\
\nonumber
&\le
n\cdot \frac1n + \sum_{j=m+1}^n \coplg(\tau \ge j)\\
\nonumber
&\le
1+\E \tau-m \\
&\le
1+\frac{4\ell_n }{\cplngg^\kn(1-\Delta_\kn)}.
\label{coupling_bound}
\end{align}

If $k_n > \ell_n$, we follow an identical line of argument with the exception that
we divide the processes into blocks of $\kn$ samples, and $\tau$ as the first time
the two processes satisfy $\Zo_{\tau}\approxkn\Zt_{\tau}$. We then bound
\begin{align*}
\coplg(\Zo_j\napproxs{\kn} \Zt_j)
=
\coplg(\Zo_j\napproxs{\kn} \Zt_j \text{ and } \tau < j)
&+
\coplg(\Zo_j\napproxs{\kn} \Zt_j \text{ and } \tau \ge j)\\
&\le \frac1n+\coplg(\tau \ge j),
\end{align*}
and finally obtain
\begin{align*}
\sum_{j=m+1}^n \coplg(\Zo_j\napproxkn \Zt_j)
&\le
n\cdot \frac1n + \sum_{j=m+1}^n \coplg(\tau \ge j)\\
&\le
1+\frac{4\kn }{\cplngg^\kn(1-\Delta_\kn)}.
\end{align*}
\eLemmap

\bCorollary
Let $\sets{V_m}_{m=0}^\ntilde$ be the Doob martingale described in Definition~\ref{dfn:stn}. Then
\[
\left|V_0-\ntilde \frac{\stn(\w)}{\stn(\Gtilde)}\right|\le 
\cB(\kn,\ell_n,\cplngg),
\]
where $\cB(\kn,\ell_n,\cplngg)$ is as defined in Definition~\ref{eq:cb}.
\Proof
We bound the value of $V_0= \E [N_{\w}|Z_0]$ by a coupling argument identical to Lemma \ref{Doob_bound}. Suppose $\sets{{\Zo_m}}$ and $\sets{{\Zt_m}}$ are coupled copies of the restriction of $\ptq$ to $\tilde{G}$, where
$\sets{{\Zo_m}}$ starts from state $Z_0$, while $\sets{{\Zt_m}}$ starts from a state chosen randomly according to the
stationary distribution of $\sets{Z_m}$. The same analysis holds and the Corollary follows using Property~\ref{prop2} in section \ref{stopping_times}.
\label{Doob_corollary}
\eCorollary
\bTheorem 
\label{stat-estim} 
Let $(\T,q(\T))$ be an unknown model in $\cM_d$. If $\{Z_m\}_{m\geq1}$ is aperiodic,
then for any $t>0$, $Y^0_{-\infty}$ and $\w\in {\tilde G}$ we have
\begin{equation*}
\ptq (|N_{\w}-\tilde{n}\frac{\stn(\w)}{\stn(\tilde{G})} \,|\geq t | Y^0_{-\infty}) 
\leq 
2\exp 
\Paren{
-\frac{(t-\cB)^2}{2\tilde{n}\cB^2}},
\end{equation*}
where $\cB=\cB(\kn,\ell_n,\cplngg)$ is as defined in~\eqref{eq:cb}.
\Proof Note that aperiodicity of the restriction $\sets{Z_m}_{m\ge1}$ of $\ptq$ to $\Gtilde$
does not require an observation of $\sets{Z_m}_{m\ge1}$. We can check for this 
property using only $\Gtilde$ as noted in Property~\ref{prop:two}.

Theorem follows by Lemma \ref{Doob_bound}, Corollary
\ref{Doob_corollary} and using Azuma's inequality.  \eTheorem 
\bRemark
The
theorem only has at most constant confidence if
$t \le {\sqrt{\ntilde}}{\cB}$. Generally speaking, 
\[
\cB
\approx
\frac{\max\sets{\kn,\ell_n}}{\cplngg^\kn(1-\Delta_{\kn})}.
\]
For at least constant confidence, the deviation of the counts of $\w$
from $\stn(\w)/\stn(\Gtilde)$ is therefore
\[
\frac{t}{\ntilde}\approx
\frac{\max\sets{\kn,\ell_n}}{\sqrt{\ntilde}\cplngg^\kn(1-\Delta_{\kn})}.
\]
For the exponential decay of dependencies, if $\ntilde=\cO(n)$, we
have both $\kn$ and $\ell_n$ of the order of $\log n$. Therefore, if $\cplngg$ 
is large enough that $\cplngg^\kn\ge n^{-\beta}$, namely
\[
\cplngg\ge \frac{\log n}{2^{\beta}\kn}
\]
for some constant $\beta<1/2$, the accuracy
to which we can estimate the stationary probability ratio
$\stn(\w)/\stn(\Gtilde)$ in Theorem~\ref{stat-estim} is  
\[
\approx
\frac{\log n}{n^{1/2-\beta}}.
\] 
\eRemark 

We conclude with a couple of remarks and conjectures.

It is to be noted that for the concentration bounds on stationary
probabilities to hold, we must have $\sum{\delta_i}<\infty$, something
that was unnecessary for the transition probabilities. At this point,
lacking a matching lower bound on deviation, we cannot say if this is
an artifact of our arguments or if this is an interesting nuance that
holds. However, we conjecture that estimating stationary probabilities
is harder---that for the case $\sum {\delta_i} $ is not finite, we may
be able to estimate only transition probabilities without ever
estimating stationary probabilities.

Finally, to actually use Theorem~\ref{stat-estim}, we further lower
bound $\cplngg$ by estimates of aggregate transition probabilities
derived from the data using Theorem~\ref{thm:aggemptht}
(or~\ref{Exponential_Die}). With the effect that the model dependent
right side in Theorem~\ref{stat-estim} is replaced by another upper
bound---potentially worse, but entirely data dependent. The new
data-only dependent upper bound holds with a reduced confidence
obtained by a union bound on the confidences of
Theorems~\ref{thm:aggemptht} (or~\ref{Exponential_Die}) and~\ref{stat-estim}. 

Note also that the accuracy to which $\cplngg$ can be estimated is the
same order of magnitude as the bound on the $\ell_1$ distance (between
the naive and aggregated parameters) given in
Theorems~\ref{thm:aggemptht} (or~\ref{Exponential_Die}). The accuracy
in Theorems~\ref{thm:aggemptht} (or~\ref{Exponential_Die}) suffices,
since we intend to use Theorem~\ref{stat-estim} when $\cplngg$ scales
$\gg \frac1{\log n}$ (as mentioned before, we like $\cplngg$ to be
$\Theta(1)$). While it is unclear if this scaling is a necessary for
$\cplngg$, we believe this could be mildly improved on.

\section{Conclusions}
We have shown how to use data generated by potentially slow mixing Markov sources to identify those states for which
naive approaches will estimate both parameters and functions related to stationary probabilities. To do so, we require
that the underlying Markov source have dependencies that are not completely arbitrary, but
die down eventually. In such cases, we show that even while the source may not have mixed (explored the state space
properly), certain properties related to contexts $\w$, namely $\tilde{q}(.|\w)$ or $\stn(\w)/\stn(\Gtilde)$ can be well
estimated, if $|\w|$ grows as $\Theta(\log n)$.  Surprisingly, we saw that it is quite possible that estimates related
to contexts $\w$ may be good, even when estimates for suffixes of $\w$ fail---the reason being
Theorem~\ref{thm:aggemptht} depends not on the source mixing, but on the dependencies dying off.
We also noted a couple of unanswered
questions in our arguments---stationary probabilities seem to be
harder to estimate, and we do not yet have a necessary condition on
how large the parameter $\cplngg$ has to be for us to expect results.

This work also uncovers a lot of open problems. The above results are sufficient to say that some estimates are
approximately accurate with high confidence. A natural, but perhaps difficult, question is whether we can give necessary
conditions on how the data must look for a given estimate to be accurate. This work also forms a cog in the growing understanding of
the information theoretic underpinnings involving estimation problems with memory. Finally, these results add to the
understanding of model classes that only admit estimators converging pointwise over the class (namely at rates that
could be arbitrarily slow depending on the underlying model), but are special in the sense that it is possible to say if the
algorithm is doing well or not.

\appendices
\section{Proof of Lemma \ref{aggregated model}}
\label{s:agg}
For all $\w \in \Ttilde$, let $F(\w)$ be the set of states $\w' \in \Ttilde$ that reach 
$\w$ in one step, i.e.,
\[
F(\w)=
\{
\w' \in \Ttilde \colon \quad \exists a \in \A \quad \text{s.t. } \w \preceq \w'a
\}.
\]
Let $\tilde{Q}$ be the transition probability matrix formed by the states of $\ptildettildeq$. 
First notice that by definition, for all $\w,\w'\in \Ttilde$, 
\begin{equation*}
\tilde{Q}(\w|\w')=
\begin{cases}
\tilde{q}(a|\w') & \text{if }   \w' \in F(\w),\\ 
0  & \text{if } \w' \notin F(\w) \\
\end{cases}
\end{equation*}
where $\tilde{q}$ is as in Definition \ref{Aggregation}. Since $\ptq$ is irreducible and aperiodic, $\ptildettildeq$ will also be irreducible and aperiodic. Thus, there is a unique stationary distribution $\tilde{\stn}$ corresponding to $\ptildettildeq$, i.e., unique solution for
\begin{equation}
\tilde{\stn}(\w)=\sum_{\w' \in \tilde{\T}} \tilde{\stn}(\w') \tilde{Q}(\w|\w') \quad \forall \w\in \tilde{\mathcal{T}}.
\label{STN_AGG}
\end{equation}
We will consider a candidate solution of the form 
\begin{equation}
\tilde{\stn}(\w)=\sum_{\V \in \T_{\w}} \stn(\V),
\label{prop_solution}
\end{equation} 
for every $\w \in \tilde{\T}$ and  show that this candidate will satisfy (\ref{STN_AGG}).
Then, the claim will follow by uniqueness of the solution. To show this, note that for all $\w\in \tilde{\T}$,
\begin{align*}
\sum_{\w' \in \tilde{\T}} \tilde{\stn}(\w') &\tilde{Q}(\w|\w')
\\&=
\sum_{\w' \in F(\w)} \tilde{\stn}(\w') \tilde{q}(a|\w') \\
&= \sum_{\w' \in F(\w)}  \bigg[ \sum_{\V \in \T_{\w'}} \stn(\V)\bigg] 
\frac{\sum_{\V \in \T_{ \w'}}{\stn(\V)q(a|\V)}}{\sum_{\V' \in \T_{ \w'}}{\stn(\V')}} \\
&\stackrel{}{=} \sum_{\w' \in F(\w)} \sum_{\V \in \T_{ \w'}}{\stn(\V) q(a|\V)}  \\
&\stackrel{(i)}{=} \sum_{\s \in \T_{\w}} \stn(\s) \\
&\stackrel{(ii)}{=}\tilde{\stn}(\w),
\end{align*}
where $(i)$ holds because
\[
\Union \limits_{\w' \in F(\w)} \T_{ \w'}=\T_{\w}, 
\]
and $(ii)$ follows from the definition of the proposed solution given in (\ref{prop_solution}).
\ignore{Note that the second statement of lemma automatically follows from the uniqueness of stationary distributions.}

\section{Proof of Proposition \ref{UB_Ent_Rate}}
\label{s:upper}
Note that for all $\w \in \Ttilde$, ${\mathcal{T}}_{\w}=\{\s \in \mathcal{T}: \w \preceq \s \} $. Since $\tilde{\mathcal{T}} \preceq \mathcal{T}$, we have
\begin{align*}
\entrTt &= \sum_{\w \in \Ttilde} \tilde{\stn}(\w) \sum_{a \in \A} \tilde{q}(a|\w) \log \frac{1}{\tilde{q}(a|\w)} \\
\ignore{&= \sum_{\w \in \Ttilde} \tilde{\stn}(\w) \sum_{a \in \A} \bigg [\frac{\sum_{\V''' \in \T_\w} \stn(\V)q(a|\V)} {\sum_{\V'''' \in \T_\w} \stn(\V')} \log \frac{\sum_{\V' \in \T_\w} \stn(\V)} {\sum_{\V'' \in \T_\w} \stn(\V)q(a|\V)}  \bigg ] \\}
&\geq \sum_{\w \in \Ttilde} \tilde{\stn}(\w) \sum_{a \in \A} \sum_{\V \in \T_\w} \bigg[ \frac{\stn(\V)}{\sum_{\V' \in \T_\w} \stn(\V')} q(a|\V) \log \frac{1}{q(a|\V)} \bigg] \\
&\aeq{(a)} \sum_{\w \in \Ttilde} \sum_{a \in \A} \sum_{\V \in \T_\w}  \stn(\V) q(a|\V) \log \frac{1}{q(a|\V)}  \\
&= \sum_{\w \in \Ttilde}  \sum_{\V \in \T_\w} \stn(\V) \sum_{a \in \A}  q(a|\V) \log \frac{1}{q(a|\V)}  \\
&= \sum_{\s \in \T} \stn(\s) \sum_{a \in \A} q(a|\s) \log \frac{1}{q(a|\s)} 
=\entrT,
\end{align*}
where the first inequality follows because
\[
\tilde{q}(a|\w)= \frac{\sum_{\V \in \T_\w} \stn(\V)q(a|\V)} {\sum_{\V' \in \T_\w} \stn(\V')},
\]
and because $g(x)=x \log \frac{1}{x}$ is concave for $x \in [0,1]$. The equality $(a)$ follows since
$
\tilde{\stn}(\w)={\sum_{\V \in \T_\w} \stn(\V)}.
$
\section{Proof of Proposition~\ref{Bonferroni}}
Wolog, let $d(i)$ be decreasing, and consider a distribution $q$ over $\mathbb{N}$ such that
\[
\sum_{j\ge i} q(j) = d(i).
\]
Let $\{X_n\}_{n\geq 1}$ be a sequence of \iid random variables distributed according to $q$ and let $E_i=\prob(X_i\geq i)$.
Therefore, $E_i$ are independent with $\mathbb{P}(E_i)=d(i)$. 
Then,
\begin{equation*}
\mathbb{P}(\bigcup_{i\geq j}E_i)=1-\prod_{i\geq j}{(1-d(i))}.  
\end{equation*}
Since 
\[
\mathbb{P}(\bigcup_{i\geq j}E_i)
\leq
\sum_{i \geq j} 
\mathbb{P}(E_i)=\sum_{i \geq j}d(i),
\]
we have
\begin{equation*}
1-\sum_{i\ge j} d(i) \le \prod_{i\ge j} (1-d(i)).
\end{equation*}
Since by assumption $0\leq d(i)\leq 1$ for all $i\ge n_0$, 
the second inequality can easily be derived by the fact that
\begin{equation*}
\prod_{i\ge j} (1-d^{2}(i))\leq 1.\eqed
\end{equation*}
\ignore{Therefore,
\begin{equation*}
\prod_{i\ge j}(1-d(i)) \le \frac1{\prod_{i\ge j} (1+d(i))}.
\end{equation*}}
\label{s:f}

\section{Proof of Proposition~\ref{die_out}}
Let $\w \in \Ttilde$ and fix $a \in \A$. Note that for $\Ttilde=\A^{k_n}$, by assumption we have for all $b',b'' \in \A$
\begin{align}
\bigg|\frac{q(a|b'\w)}{q(a|b''\w)}-1\bigg|\leq d(k_n).
\label{con1}
\end{align}
According to Lemma \ref{aggregated model}, $\tilde{q}(a|\w)$ is weighted average of $q(a|b\w), b\in \A$. Hence,
\begin{align}
\min \limits_{b \in \A}q(a|b\w) \leq \tilde{q}(a|\w) \leq \max \limits_{b \in \A}q(a|b\w).
\label{con2}
\end{align}
Combining (\ref{con1}) with (\ref{con2}) and straightforward elementary algebra shows that $\forall b \in \A$
\begin{align*}
\tilde{q}(a|\w)\big(1-d(k_n)\big) &\leq q(a|b\w) \leq \big(1+d(k_n)\big) \tilde{q}(a|\w).
\end{align*}
Proceeding inductively, for all $\s \in \T_{\w}$ we have
\begin{align*}
\bigg(\prod_{i\ge k_n} \big(1-d(i)\big) \bigg) \tilde{q}(a|\w) 
&\leq
q(a|\s) \\ 
&\leq \bigg(\prod_{i\ge k_n} \big(1+d(i)\big) \bigg) \tilde{q}(a|\w).
\end{align*}
Now, Proposition~\ref{Bonferroni} implies that
\[
\bigg (1-\sum_{i\ge k_n} d(i) \bigg ) \max_{\s\in \cT_\w} q(a|\s)
\le
\tilde{q}(a|\w)
\le
\frac{\min_{\s\in \cT_\w} q(a|\s)}
{\Paren{1-\sum_{i\ge k_n} d(i)}}.\eqed
\]
\label{s:die}

\section*{Acknowledgment}
This work was supported by NSF Grants CCF-1065632, CCF-1018984 and
EECS-1029081. The authors thank Aleksander Kav\v{c}i\'c and Rui Zhang at the
University of Hawai`i, M\={a}noa for helpful discussions, and the anonymous
reviewers for their comments and suggestions.

\bibliographystyle{IEEEtran}
\bibliography{Ref,univcod}

\begin{IEEEbiography}[{\includegraphics[scale=0.18]{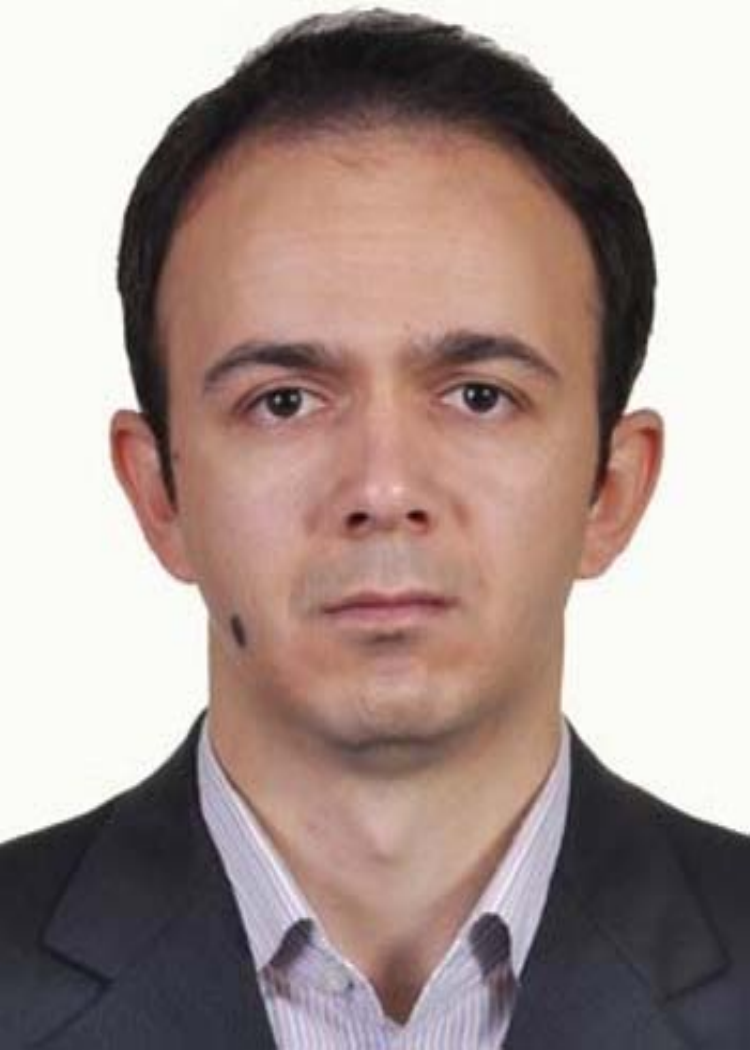}}]{Meysam Asadi}
received his B.Sc. degree
in Electrical Engineering from Razi University
of Kermanshah, Iran, in 2004, followed
by a M.Sc. degree at AmirKabir University,
Tehran, Iran, in 2007, in Electrical Engineering
and Computer Sciences. He is currently
a Ph.D. candidate in the Department of Electrical
Engineering at University of Hawai`i, M\={a}noa. His research interests cover
estimating channels with memory, estimation
in slow mixing Markov processes and detector
design for storage systems.
\end{IEEEbiography}

\begin{IEEEbiography}[{\includegraphics[scale=.12]{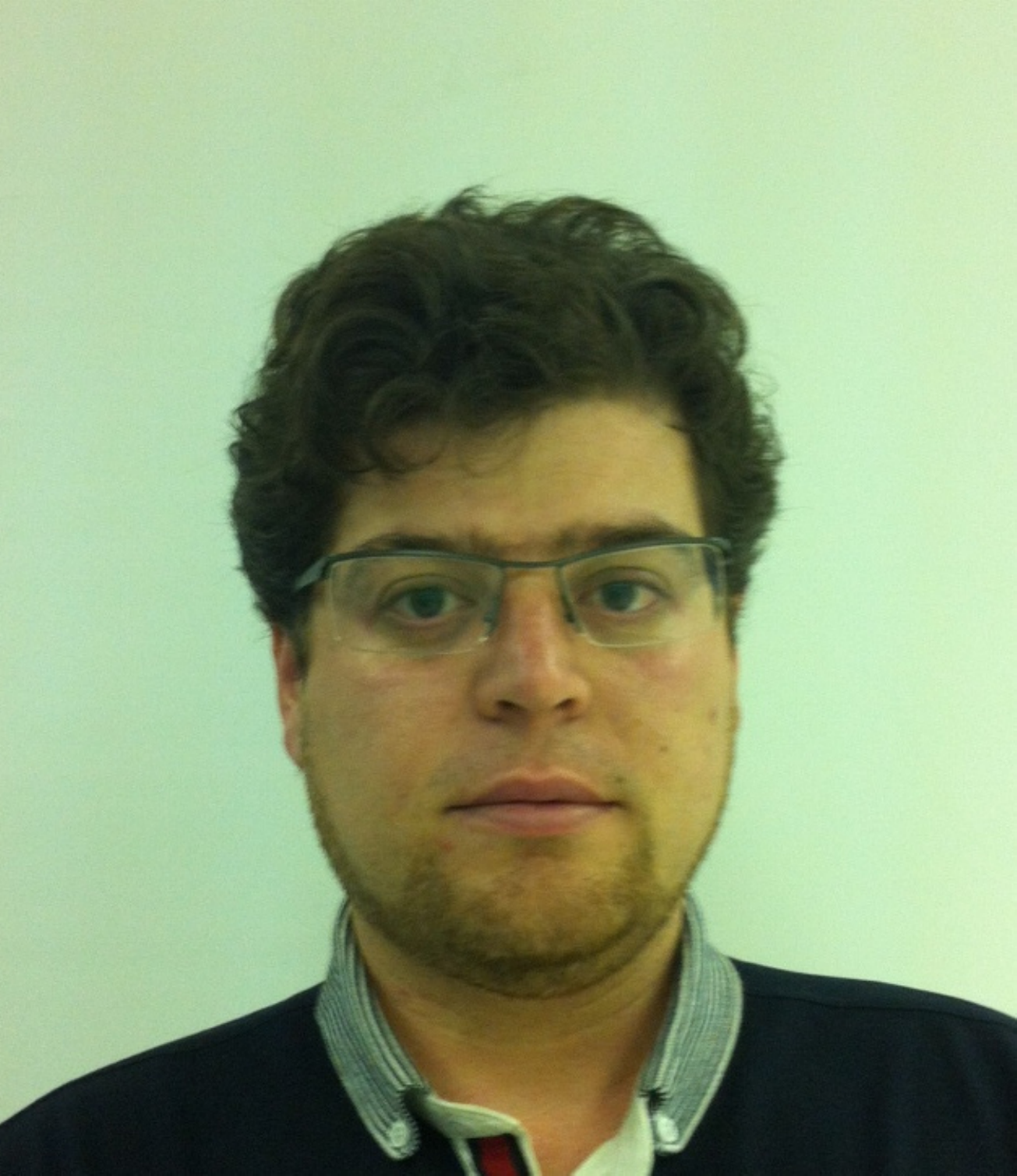}}]{Ramezan Paravi Torghabeh} was born in Mashhad, Iran, in 1982. He received Bachelor's degree from Ferdowsi University of Mashhad, Iran and Master's degree from K. N. Toosi University of Technology, Tehran, Iran both in Electrical Engineering in 2005 and 2008, respectively. He is currently a Ph.D. student in the Department of Electrical
Engineering at University of Hawai`i, M\={a}noa.

His research interests include information theory, machine learning and probabilistic methods for high dimensional problems.
He was recipient of Hawaiian Telcom Fellowship Fund in the area of telecommunications in 2011. He is a student member of the IEEE.
\end{IEEEbiography}

\begin{IEEEbiography}[{\includegraphics[scale=.08]{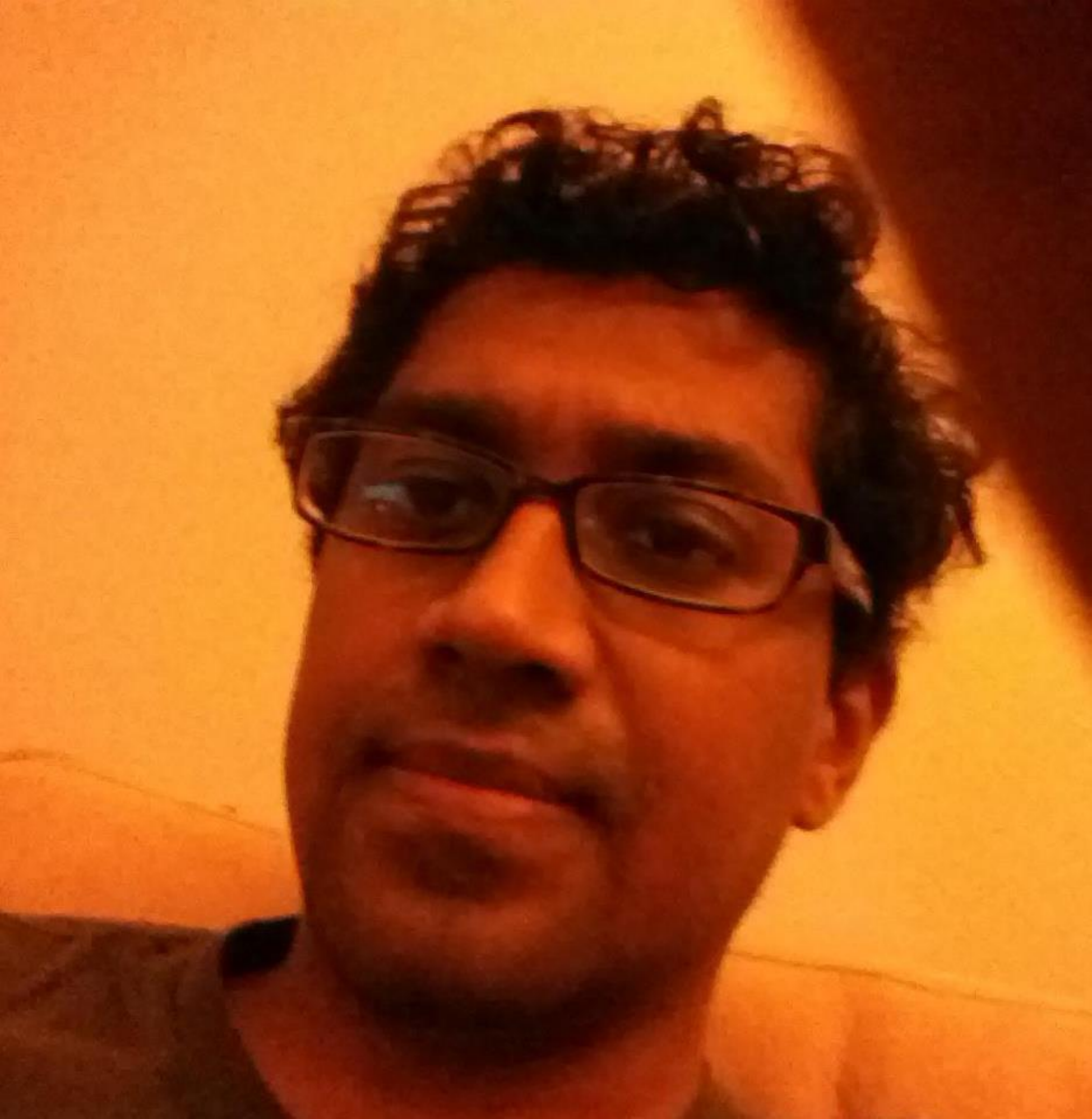}}]{Narayana Santhanam}
is an Assistant Professor at the University of
Hawaii since 2009. He obtained his B.Tech from the Indian Institute of
Technology, Chennai (then Madras) in 2000; MS and PhD from the
University of California, San Diego in 2003 and 2006 respectively.
From 2007-2008, he held a postdoctoral position at the University of
California, Berkeley.

His research interests lie in the intersection of information theory
and statistics, with a focus on the undersampled/high dimensional
regime and including applications to finance, biology, communication
and estimation theory. He is the recipient of the 2006 Information
Theory Best Paper award from the IEEE Information Theory Society along
with A. Orlitsky and J. Zhang. He has co-organized several workshops
on high dimensional statistics and "big data" problems over the last
five years, and is a member of the NSF Center for Science of
Information (CSoI), a NSF Science and Technology center.
\end{IEEEbiography}
\end{document}